\documentclass[prd,floatfix,showpacs]{revtex4}
\usepackage{epsfig}
\usepackage{array}

\newcommand{\eq}[1]{Eq.~(\ref{#1})}

\newcommand{\be}{\begin{equation}}
\newcommand{\ee}{\end{equation}}
\newcommand{\bea}{\begin{eqnarray}}
\newcommand{\eea}{\end{eqnarray}}

\newcommand{\DS}{Dyson--Schwinger }

\newcommand{\w}{\omega}

\newcommand{\al}{\alpha}
\newcommand{\ba}{\beta}
\newcommand{\ga}{\gamma}
\newcommand{\G}{\Gamma}
\newcommand{\de}{\delta}

\newcommand{\si}{\sigma}

\newcommand{\ro}{\rho}
\newcommand{\la}{\lambda}

\newcommand{\ka}{\kappa}
\newcommand{\ta}{\tau}
\newcommand{\et}{\eta}
\newcommand{\ha}{\frac{1}{2}}
\newcommand{\pd}{\partial}

\renewcommand{\th}{\theta}
\newcommand{\cd}{{\cal D}}
\newcommand{\cs}{{\cal S}}
\renewcommand{\div}{\vec{\nabla}}

\newcommand{\s}[2]{{#1}\!\cdot\!{#2}}
\newcommand{\ov}[1]{\overline{#1}}
\newcommand{\dk}[1]{\,\,\,\raisebox{-0.4ex}{\large $\bar{}$}\!\!d\,{#1}\,}
\newcommand{\dx}[1]{d^4{#1}\,}
\newcommand{\ev}[1]{<\!\!{#1}\!\!>}
\begin{document}
\title{Propagator \DS Equations of Coulomb Gauge Yang-Mills Theory Within
 the First Order Formalism}
\author{P.~Watson}
\author{H.~Reinhardt}
\affiliation{Institut f\"ur Theoretische Physik, Universit\"at T\"ubingen,
 Auf der Morgenstelle 14, D-72076 T\"ubingen, Deutschland}
\begin{abstract}
Coulomb gauge Yang-Mills theory within the first order formalism is 
considered with a view to deriving the propagator \DS equations.  The first 
order formalism is studied with special emphasis on the BRS invariance and 
it is found that there exists two forms of invariance -- invariance under 
the standard BRS transform and under a second, non-standard transform.  The 
field equations of motion and symmetries are derived explicitly and certain 
exact relations that simplify the formalism are presented.  It is shown 
that the Ward-Takahashi identity arising from invariance under the 
non-standard part of the BRS transform is guaranteed by the functional 
equations of motion.  The Feynman rules and the general decomposition of 
the two-point Green's functions are derived.  The propagator \DS equations 
are derived and certain aspects (energy independence of ghost Green's 
functions and the cancellation of energy divergences) are discussed.
\end{abstract}
\maketitle
\section{Introduction}
\setcounter{equation}{0}

Whilst there is little doubt that Quantum Chromodynamics [QCD] is the theory 
of the strong interaction, despite four decades of intense effort the 
genuine solution of the confinement puzzle and the hadron spectrum remains 
elusive.  That is not to say that no progress has been made -- our 
understanding of QCD is being steadily augmented in many ways: for example, 
with lattice Monte-Carlo techniques \cite
{Greensite:2003bk,Chandrasekharan:2004cn} and effective theories (\cite
{Bernard:2006gx} and references therein).  One way to understand the problem 
of confinement and the hadron spectrum from \emph{ab initio} principles is 
to study the \DS equations.  These equations are the central equations to 
the Lagrange formulation of a field theory.  They are in the continuum and 
embody all symmetries of the system at hand.

\DS studies of QCD in Landau gauge have enjoyed a renaissance in the last 
ten years.  A consistent picture of how the important degrees of freedom in 
the infrared (i.e., those responsible for confinement and the hadron 
spectrum) stem from the ghost sector of the theory has emerged 
\cite{vonSmekal:1997isa,vonSmekal:1997is,Watson:2001yv} and this has led to 
increasingly sophisticated calculations of QCD properties, culminating 
recently in hadron observables \cite{Fischer:2005en} and possible 
explanations of confinement (see the recent review \cite{Fischer:2006ub} for 
a discussion of this topic).  Landau gauge, in addition to having the 
appealing property of covariance, has a distinct advantage when searching 
for practical approximation schemes that allow one to extract information 
about the infrared behaviour of QCD Green's functions, namely, that the 
ghost-gluon vertex remains UV finite to all orders in perturbation theory 
\cite{Taylor:1971ff}.  For this reason, it has been possible to extract 
unambiguous information about the system \cite{Watson:2001yv}.

Despite the calculational advantages enjoyed by Landau gauge, it is perhaps 
not the best choice of gauge to study the infrared physics of QCD.  In this 
respect, Coulomb gauge is perhaps more advantageous.  
There exists a natural picture (though not a proof) of 
confinement in Coulomb gauge \cite{Zwanziger:1998ez} and phenomenological 
applications guide the way to understanding the spectrum of hadrons, for 
example in \cite{Ligterink:2003hd,Szczepaniak:2006nx} (as indeed they have 
done in Landau gauge \cite{Fischer:2006ub,Maris:2003vk}).  This is not to 
say that one choice of gauge is better than another -- it is crucial to our 
understanding of the problem that more than one gauge is considered: firstly 
because the physical observables are gauge invariant and it is a test of our 
approximations that the results respect this and secondly because whilst 
confinement is a gauge invariant reality, its mechanism may be manifested 
differently in different gauges such that we will learn far more by studying 
the different gauges.

Recently, progress has been made in studying Yang-Mills theory in Coulomb 
gauge within the Hamiltonian approach 
\cite{Szczepaniak:2001rg,
Szczepaniak:2003ve,Feuchter:2004mk,Reinhardt:2004mm}.  
Here, the advantage is that Gau\ss' law can be explicitly resolved (such 
that, in principle, gauge invariance is fully accounted for) and this 
results in an explicit expression for the static potential between color-
charges.  In 
\cite{Szczepaniak:2001rg,
Szczepaniak:2003ve,Feuchter:2004mk,Reinhardt:2004mm}, 
the Yang-Mills Schroedinger equation was solved variationally for the vacuum 
state using Gaussian type ans\"{a}tze for the wave-functional.  Minimizing 
the energy density results in a coupled set of \DS equations which have been 
solved analytically in the infrared \cite{Schleifenbaum:2006bq} and 
numerically in the entire momentum regime.  If the geometric structure of 
the space of gauge orbits reflected by the non-trivial Faddeev-Popov 
determinant is properly included \cite{Feuchter:2004mk}, one finds an 
infrared divergent gluon energy and a linear rising static quark potential 
-- both signals of confinement.  Furthermore, these confinement properties 
have been shown to be not dependent on the specific ansatz for the vacuum 
wave-functional but result from the geometric structure of the space of 
gauge orbits \cite{Reinhardt:2004mm}.  However, in spite of this success, 
one should bear in mind that an ansatz for the wave-functional is always 
required and so the approach does not {\it a priori} provide a systematic 
expansion or truncation scheme as, for example, the loop expansion scheme 
used in the common \DS approach.  A study of the \DS equations in Coulomb 
gauge will hopefully shed some light on the problem.

Given the appealing properties of Coulomb gauge, it is perhaps surprising 
that no pure \DS study exists in the literature.  However, there is a good 
reason for this: in Coulomb gauge, closed ghost-loops give rise to 
unregulated divergences -- the energy divergence problem.  It has been found 
only relatively recently how one may circumvent this problem such that a \DS 
study may be attempted \cite{Zwanziger:1998ez}.  The key lies in using the 
first order formalism.  There is not yet a complete proof that the local 
formulation of Coulomb gauge Yang-Mills theory within the first order 
formalism is renormalisable but significant progress has been made 
\cite{Zwanziger:1998ez,Baulieu:1998kx}.  There is one undesirable feature to 
the first order formalism and this is that the number of fields 
proliferates.  As will be seen in this paper, this does have serious 
implications for the \DS equations.

The purpose of the present work is to derive the \DS equations for Coulomb 
gauge Yang-Mills theory within the first order formalism.  These equations 
will form the basis for an extended program studying QCD in Coulomb gauge.  
The paper is organised as follows.  We begin in Section~2 by introducing 
Yang-Mills theory in Coulomb gauge and the first order formalism.  In 
particular, we consider the BRS invariance of the system.  Having introduced 
the first order formalism, we then motivate the reasons for considering it 
(the cancellation of the energy divergent sector and the reduction to 
physical degrees of freedom) in Section~3.  Section~4 is then concerned with 
the derivation of the equations of motion and the equations that stem from 
the BRS invariance.  There exists certain relationships that give rise to 
exact statements about the Green's functions that enter the system and these 
are detailed in Section~5.  The Feynman rules and the general decomposition 
of the two-point Green's functions are derived and discussed in Sections~6~
\&~7.  In Section~8, the \DS equations are derived in some detail and are 
discussed.  Finally, we summarize and give an outlook of future work in 
Section~9.

\section{First Order Formalism and BRS Invariance}
\setcounter{equation}{0}
Throughout this work, we work in Minkowski space and with the following 
conventions.  The metric is $g_{\mu\nu}=\mathrm{diag}(1,-\vec{1})$.  Greek 
letters ($\mu$, $\nu$, $\ldots$) denote Lorentz indices, roman subscripts 
($i$, $j$, $\ldots$) denote spatial indices and superscripts ($a$, $b$, 
$\ldots$) denote color indices.  We will sometimes also write configuration 
space coordinates ($x$, $y$, $\ldots$) as subscripts where no confusion 
arises.

The Yang-Mills action is defined as
\be
\cs_{YM}=\int\dx{x}\left[-\frac{1}{4}F_{\mu\nu}^aF^{a\mu\nu}\right]
\ee
where the (antisymmetric) field strength tensor $F$ is given in terms of the 
gauge field $A_{\mu}^a$:
\be
F_{\mu\nu}^a=\pd_{\mu}A_{\nu}^a-\pd_{\nu}A_{\mu}^a
+gf^{abc}A_{\mu}^bA_{\nu}^c.
\ee
In the above, the $f^{abc}$ are the structure constants of the $SU(N_c)$ 
group whose generators obey $\left[T^a,T^b\right]=\imath f^{abc}T^c$.  The 
Yang-Mills action is invariant under a local $SU(N_c)$ gauge transform 
characterised by the parameter $\th_x^a$:
\be
U_x=\exp{\left\{-\imath\th_x^aT^a\right\}}.
\ee
The field strength tensor can be expressed in terms of the chromo-electric 
and -magnetic fields ($\si=A^0$)
\be
\vec{E}^a=-\pd^0\vec{A}^a-\vec{\nabla}\si^a+gf^{abc}\vec{A}^b\si^c,\;\;\;\;
B_i^a=\epsilon_{ijk}\left[\nabla_jA_k^a-\ha gf^{abc}A_j^bA_k^c\right]
\ee
such that $\cs_{YM}=\int(E^2-B^2)/2$.  The electric and magnetic terms in 
the action do not mix under the gauge transform which for the gauge fields 
is written
\be
A_\mu\rightarrow A'_\mu=U_xA_\mu U_x^\dag
-\frac{\imath}{g}(\pd_\mu U_x)U_x^\dag.
\ee
Given an infinitessimal transform $U_x=1-\imath\th_x^aT^a$ the variation of 
the gauge field is
\be
\de A_{\mu}^a=-\frac{1}{g}\hat{D}_{\mu}^{ac}\th^c
\ee
where the covariant derivative in the adjoint representation is given by
\be
\hat{D}_{\mu}^{ac}=\de^{ac}\pd_{\mu}+gf^{abc}A_{\mu}^b.
\ee

Let us consider the functional integral
\be
Z=\int\cd\Phi\exp{\left\{\imath\cs_{YM}\right\}}
\ee
where $\Phi$ denotes the collection of all fields.  Since the action is 
invariant under gauge transformations, $Z$ is divergent by virtue of the 
zero mode.  To overcome this problem we use the Faddeev-Popov technique and 
introduce a gauge-fixing term along with an associated ghost term 
\cite{IZ}.  Using a Lagrange multiplier field to implement the gauge-fixing, 
in Coulomb gauge ($\s{\div}{\vec{A}}=0$) we can then write
\be
Z=\int\cd\Phi\exp{\left\{\imath\cs_{YM}+\imath\cs_{fp}\right\}},\;\;\;\;
\cs_{fp}=\int d^4x\left[-\la^a\s{\vec{\nabla}}{\vec{A}^a}
-\ov{c}^a\s{\vec{\nabla}}{\vec{D}^{ab}}c^b\right].
\ee
The new term in the action is invariant under the standard BRS transform 
whereby the infinitessimal parameter $\th^a$ is factorised into two 
Grassmann-valued components $\th^a=c^a\de\la$ where $\de\la$ is the 
infinitessimal variation (not to be confused with the colored Lagrange 
multiplier field $\la^a$).  The BRS transform of the new fields reads
\bea
\de\ov{c}^a&=&\frac{1}{g}\la^a\de\la\nonumber\\
\de c^a&=&-\ha f^{abc}c^bc^c\de\la\nonumber\\
\de\la^a&=&0.
\eea
For reasons that will become clear in the next section (expounded in 
\cite{Zwanziger:1998ez}) we convert to the first order (or phase space) 
formalism by splitting the Yang-Mills action into chromo-electric and 
-magnetic terms and introducing an auxiliary field ($\vec{\pi}$) via the 
following identity
\be
\exp{\left\{\imath\int d^4x\ha\s{\vec{E}^a}{\vec{E}^a}\right\}}
=\int\cd\vec{\pi}\exp{\left\{\imath\int d^4x\left[-\ha
\s{\vec{\pi}^a}{\vec{\pi}^a}-\s{\vec{\pi}^a}{\vec{E}^a}\right]\right\}}.
\ee
Classically, the $\vec{\pi}$-field would be the momentum conjugate to 
$\vec{A}$.  In order to maintain BRS-invariance, we require that
\be
\int d^4x\left[\s{\de\vec{\pi}^a}{\left(\vec{\pi}^a+\vec{E}^a\right)}
+\s{\vec{\pi}^a}{\de\vec{E}^a}\right]=0.
\label{eq:inv0}
\ee
Given that the variation of $\vec{E}$ under the infinitessimal gauge 
transformation is $\de\vec{E}^a=f^{abc}\vec{E}^c\th^b$, then the general 
solution to \eq{eq:inv0} is
\be
\de\vec{\pi}^a=f^{abc}\th^b\left[(1-\al)\vec{\pi}^c-\al\vec{E}^c\right]
\ee
where $\al$ is some non-colored constant, but which in general could be some 
function of position $x$.  The $\vec{\pi}$-field is split into transverse 
and longitudinal components using the identity
\bea
\mathrm{const}&=&\int\cd\phi\de\left(\s{\vec{\nabla}}{\vec{\pi}}
+\nabla^2\phi\right)\nonumber\\
&=&\int\cd\left\{\phi,\ta\right\}\exp{\left\{-\imath\int 
d^4x\ta^a\left(\s{\vec{\nabla}}{\vec{\pi}^a}+\nabla^2\phi^a\right)\right\}}.
\eea
This constant is gauge invariant and this means that the new fields $\phi$ 
and $\ta$ must transform as
\be
\de\phi^a=\s{\frac{\vec{\nabla}}{\left(-\nabla^2\right)}}{\de\vec{\pi}^a},
\;\;\de\ta^a=0.
\ee
If we make the change of variables 
$\vec{\pi}\rightarrow\vec{\pi}-\vec{\nabla}\phi$ then collecting together 
all the parts of $Z$ that contain $\vec{\pi}$, we can write
\be
Z_{\pi}=\int\cd\left\{\vec{\pi},\phi,\ta\right\}\exp{\left\{\imath\int 
d^4x\left[-\ta^a\s{\vec{\nabla}}{\vec{\pi}^a}
-\ha\s{(\vec{\pi}^a-\div\phi^a)}{(\vec{\pi}^a-\div\phi^a)}
-\s{\left(\vec{\pi}^a-\div\phi^a\right)}{\vec{E}^a}\right]\right\}}
\ee
which is now invariant under
\bea
\de\vec{E}^a&=&f^{abc}\vec{E}^c\th^b,\nonumber\\
\de\vec{\pi}^a&=&f^{abc}\th^b\left[(1-\al)\left(\vec{\pi}^c
-\vec{\nabla}\phi^c\right)-\al\vec{E}^c\right]+\vec{\nabla}\de\phi^a,
\nonumber\\
\de\phi^a&=&f^{abc}\left\{\s{\frac{\vec{\nabla}}{\left(-\nabla^2\right)}}
{\left[(1-\al)\left(\vec{\pi}^c-\vec{\nabla}\phi^c\right)
-\al\vec{E}^c\right]}\th^b\right\},\nonumber\\
\de\ta^a&=&0.
\eea
We notice that the parts of the transform that are proportional to $\al$ are 
independent of the rest of the BRS transform and can thus be regarded as a 
separate invariance.  In particular, since it is independent of the 
Faddeev--Popov components, then we may regard it quite generally as a local 
transform parameterised by $\th^a$.  This new invariance stems from the 
arbitrariness in introducing the $\vec{\pi}$-field.  If we expand the 
chromo-electric field into its component form then in summary we can write 
our full functional integral as
\be
Z=\int\cd\Phi\exp{\left\{\imath\cs_B+\imath\cs_{fp}+\imath\cs_{\pi}\right\}}
\label{eq:func}
\ee
with
\bea
\cs_B&=&\int d^4x\left[-\ha\s{\vec{B}^a}{\vec{B}^a}\right],\nonumber\\
\cs_{fp}&=&\int d^4x\left[-\la^a\s{\vec{\nabla}}{\vec{A}^a}
-\ov{c}^a\s{\vec{\nabla}}{\vec{D}^{ab}}c^b\right],\nonumber\\
\cs_{\pi}&=&\int d^4x\left[-\ta^a\s{\vec{\nabla}}{\vec{\pi}^a}
-\ha\s{(\vec{\pi}^a-\div\phi^a)}{(\vec{\pi}^a-\div\phi^a)}
+\s{(\vec{\pi}^a-\div\phi^a)}{\left(\pd^0\vec{A}^a+\vec{D}^{ab}\si^b
\right)}\right]
\label{eq:act}
\eea
and which is invariant under \emph{two} sets of transforms: the BRS
\bea
\de\vec{A}^a&=&\frac{1}{g}\vec{D}^{ac}c^c\de\la,\nonumber\\
\de\si^a&=&-\frac{1}{g}D^{0ac}c^c\de\la,\nonumber\\
\de\ov{c}^a&=&\frac{1}{g}\la^a\de\la,\nonumber\\
\de c^a&=&-\ha f^{abc}c^bc^c\de\la,\nonumber\\
\de\vec{\pi}^a&=&f^{abc}c^b\de\la\left(\vec{\pi}^c
-\vec{\nabla}\phi^c\right)+\vec{\nabla}\de\phi^a,\nonumber\\
\de\phi^a&=&f^{abc}\left\{\s{\frac{\vec{\nabla}}{\left(-\nabla^2\right)}}
{\left(\vec{\pi}^c-\vec{\nabla}\phi^c\right)}c^b\de\la\right\},\nonumber\\
\de\la^a&=&0,\nonumber\\
\de\ta^a&=&0,
\label{eq:brstrans}
\eea
and the new transform, which we denote the $\al$-transform,
\bea
\de\vec{\pi}^a&=&f^{abc}\th^b\left(\vec{\pi}^c-\vec{\nabla}\phi^c
-\pd^0\vec{A}-\vec{D}^{cd}\si^d\right)+\vec{\nabla}\de\phi^a,\nonumber\\
\de\phi^a&=&f^{abc}\left\{\s{\frac{\vec{\nabla}}{\left(-\nabla^2\right)}}
{\left(\vec{\pi}^c-\vec{\nabla}\phi^c-\pd^0\vec{A}^c-\vec{D}^{cd}\si^d
\right)}\th^b\right\}
\label{eq:altrans}
\eea
(all other fields being unchanged).  It useful for later to denote the 
combination of fields and differential operators occuring in \eq{eq:altrans} 
as
\be
\vec{X}^c=\vec{\pi}^c-\vec{\nabla}\phi^c-\pd^0\vec{A}-\vec{D}^{cd}\si^d.
\label{eq:X}
\ee

\section{Formal Reduction to ``Physical" Degrees of Freedom}
\setcounter{equation}{0}
There are two factors that motivate our use of the first order formalism.  
The first lies in the ability, albeit formally, to reduce the functional 
integral previously considered (and hence the generating functional) to 
``physical" degrees of freedom \cite{Zwanziger:1998ez}.  These are the 
transverse gluon and transverse $\vec{\pi}$ fields which in classical terms 
would be the configuration variables and their momentum conjugates.  We keep 
the term ``physical" in quotation marks because it is realised that in 
Yang-Mills theory, the true physical objects would be the color singlet 
glueballs, their observables being the mass spectrum and the decay widths.  
The second factor concerns the well-known energy divergence problem of 
Coulomb gauge QCD \cite{Heinrich:1999ak,Andrasi:2005xu,Doust:1987yd}.  In 
Coulomb gauge, the Faddeev-Popov operator involves only spatial derivatives 
and the spatial components of the gauge fields, but these fields are 
themselves dependent on the spacetime position.  This leads to the ghost 
propagator and ghost-gluon vertex being independent of the energy whereas 
loops involving pure ghost components are integrated over both 3-momentum 
{\it and} energy which gives an ill-defined integration.  In the usual, 
second order, formulation of the theory these energy divergences do in 
principle cancel order by order in perturbation theory (tested up to 
two-loops \cite{Heinrich:1999ak}) but this cancellation is difficult to 
isolate.  Within the first order formalism the cancellation is made manifest 
such that the problem of ill-defined integrals can be circumvented.

Given the functional integral, \eq{eq:func}, and the action, \eq{eq:act}, we 
rewrite the Lagrange multiplier terms as $\de$-function constraints and the 
ghost terms as the original Faddeev-Popov determinant.  Since the 
$\de$-function constraints are now exact we can automatically eliminate any 
$\s{\div}{\vec{A}}$ and $\s{\div}{\vec{\pi}}$ terms in the action.  This is 
clearly at the expense of a local formulation and the BRS invariance of the 
theory is no longer manifest.  The functional integral is now
\be
Z=\int\cd\Phi\mathrm{Det}\left[-\s{\vec{\nabla}}{\vec{D}}\de^4(x-y)\right]
\de\left(\s{\div}{\vec{A}}\right)\de\left(\s{\div}{\vec{\pi}}\right)
\exp{\left\{\imath\cs\right\}}
\ee
with
\be
\cs=\int d^4x\left[-\ha\s{\vec{B}^a}{\vec{B}^a}
-\ha\s{\vec{\pi}^a}{\vec{\pi}^a}+\ha\phi^a\nabla^2\phi^a
+\s{\vec{\pi}^a}{\pd^0\vec{A}^a}+\si^a\left(\s{\div}{\vec{D}^{ab}}\phi^b
+g\hat{\ro}^a\right)\right].
\ee
where we have defined an effective charge 
$\hat{\ro}^a=f^{ade}\s{\vec{A}^d}{\vec{\pi}^e}$.  The integral over $\si$ 
can also be written as a $\de$-function constraint and is the implementation 
of the chromo-dynamical equivalent of Gau\ss' law giving
\be
Z=\int\cd\Phi\mathrm{Det}\left[-\s{\vec{\nabla}}{\vec{D}}\de^4(x-y)\right]
\de\left(\s{\div}{\vec{A}}\right)\de\left(\s{\div}{\vec{\pi}}\right)
\de\left(-\s{\div}{\vec{D}^{ab}}\phi^b-g\hat{\ro}^a\right)
\exp{\left\{\imath\cs\right\}}
\ee
with
\be
\cs=\int d^4x\left[-\ha\s{\vec{B}^a}{\vec{B}^a}
-\ha\s{\vec{\pi}^a}{\vec{\pi}^a}+\ha\phi^a\nabla^2\phi^a
+\s{\vec{\pi}^a}{\pd^0\vec{A}^a}\right].
\ee
Let us define the inverse Faddeev-Popov operator $M$:
\be
\left[-\s{\div}{\vec{D}^{ab}}\right]M^{bc}=\de^{ac}.
\ee
With this definition we can factorise the Gau{\ss} law $\de$-function 
constraint as
\be
\de\left(-\s{\div}{\vec{D}^{ab}}\phi^b-g\hat{\ro}^a\right)
=\mathrm{Det}\left[-\s{\vec{\nabla}}{\vec{D}}\de^4(x-y)\right]^{-1}
\de\left(\phi^a-M^{ab}g\hat{\ro}^b\right).
\ee
Crucially, the inverse functional determinant cancels the original 
Faddeev-Popov determinant, leaving us with
\be
Z=\int\cd\Phi\de\left(\s{\div}{\vec{A}}\right)
\de\left(\s{\div}{\vec{\pi}}\right)\de\left(\phi^a-M^{ab}g\hat{\ro}^b\right)
\exp{\left\{\imath\cs\right\}}.
\ee
We now use the $\de$-function constraint to eliminate the $\phi$-field.  
Recognising the Hermitian nature of the inverse Faddeev-Popov operator $M$ 
we can reorder the operators in the action to give us
\be
Z=\int\cd\Phi\de\left(\s{\div}{\vec{A}}\right)
\de\left(\s{\div}{\vec{\pi}}\right)\exp{\left\{\imath\cs\right\}}
\ee
with
\be
\cs=\int d^4x\left[-\ha\s{\vec{B}^a}{\vec{B}^a}
-\ha\s{\vec{\pi}^a}{\vec{\pi}^a}
-\ha g\hat{\ro}^bM^{ba}(-\nabla^2)M^{ac}g\hat{\ro}^c
+\s{\vec{\pi}^a}{\pd^0\vec{A}^a}\right].
\ee
The above action is our desired form, with only transverse $\vec{A}$ and 
$\vec{\pi}$ fields present.  All other fields, especially those responsible 
for the Faddeev-Popov determinant (i.e., the certainly unphysical ghosts) 
have been formally eliminated.  However, the appearance of the functional 
$\de$-functions and the inverse Faddeev-Popov operator $M$ have led to a 
non-local formalism.  It is not known how to use forms such as the above in 
calculational schemes.  The issue of renormalisability is certainly unclear 
and one does not have a Ward identity in the usual sense.

The non-local nature of the above result may not lend itself to 
calculational devices but does serve as a guide to the local formulation.  
In particular, it is evident that decomposition of degrees of freedom, both 
physical and unphysical, inherent to the first order formalism leads more 
naturally to the cancellation of the unphysical components in the 
description of physical phenomena than perhaps other choices such as Landau 
gauge.  The task ahead is to identify, within the local formulation, how 
these cancellations arise and to ensure that approximation schemes respect 
such cancellations.  For example, the cancellation of the Faddeev-Popov 
determinant and the appearance of the inverse Faddeev-Popov operator should 
lead to the separation of the physical gluon dynamics contained within the 
ghost sector and the unphysical ghosts themselves, i.e., the unphysical 
ghost loop of the gluon polarisation should be cancelled whilst another loop 
containing only physical information will take its place.  Also, it should 
be evident that the energy divergences associated with ghost loops are 
explicitly cancelled such that ill-defined integrals do not occur.

\section{Field Equations of Motion and Continuous Symmetries}
\setcounter{equation}{0}
The generating functional of the theory is given by our previously 
considered functional integral in the presence of sources.  Explicitly, 
given the action, \eq{eq:act}, we have
\be
Z[J]=\int\cd\Phi\exp{\left\{\imath\cs_B+\imath\cs_{fp}
+\imath\cs_{\pi}+\imath\cs_s\right\}}
\label{eq:gen0}
\ee
with sources defined by
\be
\cs_s=\int\dx{x}\left[\ro^a\si^a+\s{\vec{J}^a}{\vec{A}^a}+\ov{c}^a\et^a
+\ov{\et}^ac^a+\ka^a\phi^a+\s{\vec{K}^a}{\vec{\pi}^a}+\xi_\la^a\la^a
+\xi_\ta^a\ta^a\right].
\label{eq:source0}
\ee
It is useful to introduce a compact notation for the sources and fields and 
we denote a generic field $\Phi_\al$ with source $J_\al$ such that the index 
$\al$ stands for all attributes of the field in question (including its 
type) such that, for instance, we could write
\be
\cs_s=J_\al\Phi_\al
\ee
where summation over all discrete indices and integration over all 
continuous arguments is implicitly understood.

The field equations of motion are derived from the observation that the 
integral of a total derivative vanishes up to boundary terms.  The boundary 
terms vanish but this is not so trivial in the light of the Gribov problem.  
Perturbatively (expanding around the free-field), there are certainly no 
boundary terms to be considered, however, nonperturbatively the presence of 
so-called Gribov copies does complicate the picture somewhat.

In \cite{Gribov:1977wm}, Gribov showed that the Faddeev-Popov technique does 
not uniquely fix the gauge and showed that even after gauge-fixing there are 
(physically equivalent) gauge configurations related by finite gauge 
transforms still present.  It was proposed to restrict the space of gauge 
field configurations ($A$) to the so-called Gribov region $\Omega$ defined by
\be
\Omega\equiv\left\{A:\s{\div}{\vec{A}}=0;-\s{\div}{\vec{D}}\geq0\right\}.
\ee
$\Omega$ is a region where the Coulomb gauge condition holds and 
furthermore, the eigenvalues of the Faddeev-Popov operator are all 
positive.  It contains the element $\vec{A}=0$ and is bounded in every 
direction \cite{Zwanziger:1998yf}.  However, as explained in 
\cite{Zwanziger:2003cf} and references therein, the Gribov region is not 
entirely free of Gribov copies and one should more correctly consider the 
fundamental modular region $\Lambda$ which is defined as the region free of 
Gribov copies.  It turns out though that the functional integral is 
dominated by configurations on the common boundary of $\Lambda$ and $\Omega$ 
so that, in practise, restriction to the Gribov region is sufficient.

Given that non-trivial boundary conditions are being imposed (i.e., 
restricting to the Gribov region $\Omega$), the question of the boundary 
terms in the derivation of the field equations of motion now becomes 
extremely relevant.  However, by definition on the boundary of $\Omega$, the 
Faddeev-Popov determinant {\it vanishes} such that the boundary terms are 
identically zero.  The form of the field equations of motion is therefore 
equivalent as if we had extended the integration region to the full 
configuration space \cite{Zwanziger:2003cf}.

Writing $\cs=\cs_B+\cs_{fp}+\cs_{\pi}$ we have that
\be
0=\int\cd\Phi\frac{\de}{\de\imath\Phi_\al}
\exp{\left\{\imath\cs+\imath\cs_s\right\}}
=\int\cd\Phi\left\{\frac{\de\cs}{\de\Phi_\al}
+\frac{\de\cs_s}{\de\Phi_\al}\right\}
\exp{\left\{\imath\cs+\imath\cs_s\right\}}
\ee
and so, taking advantage of the linearity in the fields of the source term 
of the action we have
\be
J_{\al}Z=-\int\cd\Phi\left\{\frac{\de\cs}{\de\Phi_\al}\right\}
\exp{\left\{\imath\cs+\imath\cs_s\right\}}.
\label{eq:eom}
\ee
We use the convention that all Grassmann-valued derivatives are 
left-derivatives and so in the above there will be an additional minus sign 
on the left-hand side when $\al$ refers to derivatives with respect to 
either the $c$-field or the $\et$-source.  The explicit form of the various 
field equations of motion are given in Appendix~\ref{app:eom}.

Continuous transforms, under which the action is invariant, can be regarded 
as changes of variable and providing that the Jacobian is trivial, one is 
left with an equation relating the variations of the source terms in the 
action.  We consider the two invariances derived explicitly in the previous 
section and that the Jacobian factors are trivial is shown in 
Appendix~\ref{app:jac}.  In the case of the BRS transform, \eq{eq:brstrans}, 
we have
\bea
0&=&\int\cd\Phi\frac{\de}{\de\imath\de\la}
\exp{\left\{\imath\cs+\imath\cs_s+\imath\de\cs_s\right\}}_{\de\la=0}
\nonumber\\
&=&\int\cd\Phi\int\dx{x}\left\{-\frac{1}{g}\ro^aD^{0ab}c^b
+\frac{1}{g}\s{\vec{J}^a}{\vec{D}^{ab}}c^b-\frac{1}{g}\la^a\et^a
-\ha f^{abc}\ov{\et}^ac^bc^c
\right.\nonumber\\&&\left.
+f^{abc}c^b\s{(\vec{\pi}^c-\div\phi^c)}{\left[\vec{K}^a-\frac{\div}{
(-\nabla^2)}(\ka^a-\s{\div}{\vec{K}^a})\right]}\right\}
\exp{\left\{\imath\cs+\imath\cs_s\right\}}.
\eea
Notice that the infinitessimal variation $\de\la$ that parameterises the BRS 
transform is a global quantity, leading to the overall integral over $x$.  
The equation for the $\al$-transform is:
\bea
0&=&\int\cd\Phi\frac{\de}{\de\imath\th_x^a}
\exp{\left\{\imath\cs+\imath\cs_s+\imath\de\cs_s\right\}}_{\th=0}\nonumber\\
&=&\int\cd\Phi f^{abc}\s{\vec{X}_x^c}{\left[\vec{K}_x^a-\frac{\div_x}{
(-\nabla_x^2)}(\ka_x^a-\s{\div_x}{\vec{K}_x^a})\right]}
\exp{\left\{\imath\cs+\imath\cs_s\right\}}
\label{eq:alinv}
\eea
where $\vec{X}$ is given by \eq{eq:X}.  Constraints imposed by the discrete 
symmetries of time-reversal and parity will be discussed later.

The above equations of motion and symmetries refer to functional derivatives 
of the full generating functional.  In practise we are concerned with 
connected two-point (propagator) and one-particle irreducible $n$-point 
(proper) Green's functions since these comprise the least sophisticated 
common building blocks from which all other amplitudes may be constructed.  
The generating functional of connected Green's functions is $W[J]$ where
\be
Z[J]=e^{W[J]}.
\ee
We introduce a bracket notation for functional derivatives of $W$ such that
\be
\ev{\imath J_1}=\frac{\de W}{\de\imath J_1}.
\ee
The classical field $\Phi_\al$ is defined as
\be
\Phi_\al=\frac{1}{Z}\int\cd\Phi\Phi_\al
\exp{\left\{\imath\cs+\imath\cs_s\right\}}
=\frac{1}{Z}\frac{\de Z}{\de\imath J_\al}
\ee
(the classical field is distinct from the quantum fields which are 
functionally integrated over, but for convenience we use the same 
notation).  The generating functional of proper Green's functions is the 
effective action $\G$, which is a function of the classical fields and is 
defined via a Legendre transform of $W$:
\be
\G[\Phi]=W[J]-\imath J_\al\Phi_\al.
\ee
We use the same bracket notation to denote derivatives of $\G$ with respect 
to fields -- no confusion arises since we never mix derivatives with respect 
to sources and fields.

Let us now present the equations of motion in terms of proper functions 
(from which we will derive the \DS equations).  Using the  equations of 
motion listed in Appendix~\ref{app:eom} we have the following equations:
\begin{itemize}
\item
$\si$-based.  This is the functional form of Gau\ss' law.
\bea
\ev{\imath\si_x^a}-\ev{\imath\ta_x^a}&=&\nabla_x^2\phi_x^a
+gf^{abc}A_{ix}^b\pi_{ix}^c-gf^{abc}A_{ix}^b\nabla_{ix}\phi_x^c
+gf^{abc}\ev{\imath J_{ix}^b\imath K_{ix}^c}
\nonumber\\&&
-gf^{abc}\int d^4y\de(x-y)\nabla_{ix}\ev{\imath J_{iy}^b\imath\ka_x^c}.
\label{eq:sidse0}
\eea
Note that we have implicitly used the $\ta$ equation of motion, 
\eq{eq:tadse1}, in order to eliminate terms involving $\s{\div}{\vec{\pi}}$ 
in favour of the source $\xi_{\ta}$.
\item
$\vec{A}$-based.  We write this in such a way as to factorise the functional 
derivatives and the kinematical factors.  The equation reads:
\bea
\ev{\imath A_{ix}^a}&=&\nabla_{ix}\la_x^a-\pd_x^0\pi_{ix}^a
+\pd_x^0\nabla_{ix}\phi_x^a+\left[\de_{ij}\nabla_x^2
-\nabla_{ix}\nabla_{jx}\right]A_{jx}^a
\nonumber\\&&
+gf^{abc}\int\dx{y}\dx{z}\de(y-x)\de(z-x)\left[\nabla_{iz}\ov{c}_z^bc_y^c
+\pi_{iz}^b\si_y^c-\nabla_{iz}\phi_z^b\si_y^c\right]
\nonumber\\&&
+gf^{abc}\int\dx{y}\dx{z}\de(y-x)\de(z-x)
\left[\nabla_{iz}\ev{\imath\ov{\et}_y^c\imath\et_z^b}
+\ev{\imath K_{iz}^b\imath\ro_y^c}
-\nabla_{iz}\ev{\imath\ka_z^b\imath\ro_y^c}\right]
\nonumber\\&&
+gf^{abc}\int\dx{y}\dx{z}\de(y-x)\de(z-x)\left\{\de_{jk}\nabla_{iz}
+2\de_{ij}\nabla_{ky}-\de_{ik}\nabla_{jy}\right\}
\left[\ev{\imath J_{jy}^b\imath J_{kz}^c}+A_{jy}^bA_{kz}^c\right]
\nonumber\\&&
-\frac{1}{4}g^2f^{fbc}f^{fde}\de_{jk}\de_{li}
\left[\de^{cg}\de^{eh}(\de^{ab}\de^{di}+\de^{ad}\de^{bi})
+\de^{bg}\de^{dh}(\de^{ac}\de^{ie}+\de^{ae}\de^{ic})\right]\times
\nonumber\\&&
\left[\ev{\imath J_{jx}^g\imath J_{kx}^h\imath J_{lx}^i}
+A_{jx}^g\ev{\imath J_{kx}^h\imath J_{lx}^i}
+A_{kx}^h\ev{\imath J_{jx}^g\imath J_{lx}^i}
+A_{lx}^i\ev{\imath J_{jx}^g\imath J_{kx}^h}+A_{jx}^gA_{kx}^hA_{lx}^i\right].
\label{eq:adse0}
\eea
\item
ghost-based.  The ghost and the antighost equations provide the same 
information.  The two fields are complimentary and derivatives must come in 
pairs if the expression is to survive when sources are set to zero.  The 
antighost equation is
\be
\ev{\imath\ov{c}_x^a}=-\nabla_x^2c_x^a
-gf^{abc}\nabla_{ix}
\left[\ev{\imath\ov{\et}_x^b\imath J_{ix}^c}+c_x^bA_{ix}^a\right].
\label{eq:ghost0}
\ee
\item
$\vec{\pi}$-based.
\be
\ev{\imath\pi_{ix}^a}=\nabla_{ix}\ta_x^a-\pi_{ix}^a+\nabla_{ix}\phi_x^a
+\pd_x^0A_{ix}^a+\nabla_{ix}\si_x^a
+gf^{abc}\left[\ev{\imath\ro_x^b\imath J_{ix}^c}+\si_x^bA_{ix}^c\right].
\label{eq:pidse0}
\ee
\item
$\phi$-based.  We notice that the interaction terms in the equation of 
motion for the $\phi$-field, \eq{eq:phdse1} are, up to a derivative, 
\emph{identical} to those of the $\vec{\pi}$-based equation, \eq{eq:pidse1}.  
This arises since $-\div\phi$ is nothing more than the longitudinal part of 
$\vec{\pi}$ and means that there is a redundancy in the formalism that can 
be exploited to simplify proceedings.  We can write
\be
\left(\s{\div_x}{\vec{K}_x^a}-\ka_x^a\right)Z
=-\int\cd\Phi\nabla_x^2\ta_x^a\exp{\left\{\imath\cs+\imath\cs_s\right\}},
\label{eq:phidse1}
\ee
from which it follows that
\bea
\label{eq:phidse}\s{\div_x}{\vec{K}_x^a}-\ka_x^a&=&
-\nabla_x^2\ev{\imath\xi_{\ta x}^a},\\
\label{eq:phidse0}\ev{\imath\phi_x^a}
-\s{\div_x}{\ev{\imath\vec{\pi}_x^a}}&=&-\nabla_x^2\ta_x^a.
\eea
\item
$\la$- and $\ta$-based.  It will be useful to have these equations written 
in terms of both connected and proper Green's functions.
\bea
\label{eq:ladse}\xi_{\la x}^a=\s{\div_x}{\ev{\imath\vec{J}_x^a}},&
\;\;\;\;&\ev{\imath\la_x^a}=-\s{\div_x}{\vec{A}_x^a},\\
\label{eq:tadse}\xi_{\ta x}^a=\s{\div_x}{\ev{\imath\vec{K}_x^a}},&
\;\;\;\;&\ev{\imath\ta_x^a}=-\s{\div_x}{\vec{\pi}_x^a}.
\eea
\end{itemize}
The BRS transform gives rise to the following equation (the Ward--Takahashi 
identity):
\bea
0&=&\int d^4x\left\{\frac{1}{g}(\pd_x^0\ro_x^a)\ev{\imath\ov{\et}_x^a}
-f^{acb}\ro_x^a\left[\ev{\imath\ro_x^c\imath\ov{\et}_x^b}
+\ev{\imath\ro_x^c}\ev{\imath\ov{\et}_x^b}\right]
-\frac{1}{g}(\nabla_{ix}J_{ix}^a)\ev{\imath\ov{\et}_x^a}
\right.\nonumber\\&&
-f^{acb}J_{ix}^a\left[\ev{\imath J_{ix}^c\imath\ov{\et}_x^b}
+\ev{\imath J_{ix}^c}\ev{\imath\ov{\et}_x^b}\right]
-\frac{1}{g}\et_x^a\ev{\imath\xi_{\la x}^a}
-\ha f^{abc}\ov{\et}_x^a\left[\ev{\imath\ov{\et}_x^b\imath\ov{\et}_x^c}
+\ev{\imath\ov{\et}_x^b}\ev{\imath\ov{\et}_x^c}\right]
\nonumber\\&&
+f^{abc}\left[K_{ix}^a-\frac{\nabla_{ix}}{(-\nabla_x^2)}
(\ka_x^a-\nabla_{jx}K_{jx}^a)\right]\times
\nonumber\\&&\left.
\left[\ev{\imath K_{ix}^c\imath\ov{\et}_x^b}
-\int\dx{y}\de(x-y)\nabla_{ix}\ev{\imath\ka_x^c\imath\ov{\et}_y^b}
+\ev{\imath K_{ix}^c}\ev{\imath\ov{\et}_x^b}
-\ev{\imath\ov{\et}_x^b}\nabla_{ix}\ev{\imath\ka_x^c}\right]\right\}.
\label{eq:stid0}
\eea
We consider for now only the form of the equation relating connected Green's 
functions.  As will be seen in the next section, it will not be necessary to 
consider the equation generated by the invariance under the $\al$-transform.

\section{Exact Relations for Green's Functions}
\setcounter{equation}{0}
Given the set of `master' field equations of motion and symmetries, it is 
pertinent to find out if any of the constraints can be combined to give 
unambiguous information about the eventual Green's functions of the theory.  
We find that such simplifications do in fact exist.

Let us start by discussing the functional equation generated by 
$\al$-invariance, \eq{eq:alinv}.  It is not necessary here to consider 
functional derivatives of either the generating functional of connected 
Green's functions ($W$) or the effective action ($\G$) since the derivation 
applies to the functional integrals directly.  From Appendix~\ref{app:eom}, 
the $\vec{\pi}$-based field equation of motion, \eq{eq:pidse1} is
\be
K_{ix}^aZ[J]=-\int\cd\Phi\left\{\nabla_{ix}\ta_x^a-X_{ix}^a\right\}
\exp{\left\{\imath\cs+\imath\cs_s\right\}},\label{eq:pidse2}
\ee
Using \eq{eq:pidse2} we can rewrite \eq{eq:alinv} as
\be
0=f^{abc}\int\cd\Phi\s{\left[\vec{K}_x^a-\frac{\div_x}{(-\nabla_x^2)}
(\ka_x^a-\s{\div_x}{\vec{K}_x^a})\right]}{\left[\vec{K}_x^c
+\div_x\ta_x^c\right]}\exp{\left\{\imath\cs+\imath\cs_s\right\}}.
\label{eq:alinv1}
\ee
Since $f^{abc}$ is antisymmetric and noting \eq{eq:phidse1}, the above is 
now an almost trivial identity.  We have thus shown that the 
$\vec{\pi}$- and $\phi$-based equations of motion, \eq{eq:pidse2} and 
\eq{eq:phidse1}, guarantee that $\al$-invariance is respected.  Conversely, 
approximations to the equations of motion will destroy the symmetry.  This 
is a concrete example of a general feature of any physical field theory -- 
the full solutions of the field equations of motion (and the subsequent 
functional derivatives which comprise the \DS equations) contain all the 
information given by the symmetry considerations.  In this case, we have the 
ambiguity associated with introducing the $\vec{\pi}$-field and assigning 
its properties under the BRS transform encoded within the invariance under 
the $\al$-transform and the field equations of motion are `aware' of this.  
What is unusual about this, however, is that the equivalence of the field 
equations of motion and the equations generated by invariance under a 
symmetry is invariably impossible to show (except order by order in 
perturbation theory) -- full gauge invariance being the archetypal example.

Let us now continue the discussion by considering those equations of motion 
which do not contain interaction terms.  In the absence of interactions, the 
solutions to these equations can be written down without difficulty.  In 
terms of connected Green's function, the only non-zero functional derivative 
of the $\la$-equation, \eq{eq:ladse} is
\be
\s{\div_x}{\ev{\imath\xi_{\la y}^b\imath\vec{J}_x^a}}
=-\imath\de^{ba}\de(y-x),
\label{eq:ladse2}
\ee
the right-hand side vanishing for all other derivatives.  Separating the 
configuration space arguments and setting our conventions for the Fourier 
transform, we have for a general two-point function (connected or proper) 
which obeys translational invariance:
\be
\ev{\imath J_{\al}(y)\imath J_{\ba}(x)}
=\ev{\imath J_{\al}(y-x)\imath J_{\ba}(0)}
=\int\dk{k}W_{\al\ba}(k)e^{-\imath k\cdot(y-x)}
\ee
where $\dk{k}=d^4k/(2\pi)^4$ and it is implicitly understood that the 
relevant prescription to avoid integration over poles is present such that 
the analytic continuation to Euclidean space may be performed.  We can 
immediately write down the functional derivatives of $\ev{\imath J_{ix}^q}$ 
using \eq{eq:ladse}:
\bea
\ev{\imath J_{jy}^b\imath J_{ix}^a}&=&
\int\dk{k}W_{AAji}^{ba}(k)t_{ji}(\vec{k})e^{-\imath k\cdot(y-x)},\nonumber\\
\ev{\imath K_{jy}^b\imath J_{ix}^a}&=&
\int\dk{k}W_{\pi Aji}^{ba}(k)t_{ji}(\vec{k})e^{-\imath k\cdot(y-x)},
\nonumber\\
\ev{\imath \xi_{\la y}^b\imath J_{ix}^a}&=&
\int\dk{k}\de^{ba}\frac{k_i}{\vec{k}^2}e^{-\imath k\cdot(y-x)},\nonumber\\
\ev{\imath \xi_{\ta y}^b\imath J_{ix}^a}&=&
\ev{\imath \ro_y^b\imath J_{ix}^a}=\ev{\imath \ka_y^b\imath J_{ix}^a}=0.
\eea
Similarly, for the $\ta$-equation, \eq{eq:tadse}, we have
\bea
\ev{\imath J_{jy}^b\imath K_{ix}^a}&=&
\int\dk{k}W_{A\pi ji}^{ba}(k)t_{ji}(\vec{k})e^{-\imath k\cdot(y-x)}
,\nonumber\\
\ev{\imath K_{jy}^b\imath K_{ix}^a}&=&
\int\dk{k}W_{\pi\pi ji}^{ba}(k)t_{ji}(\vec{k})e^{-\imath k\cdot(y-x)},
\nonumber\\
\ev{\imath \xi_{\ta y}^b\imath K_{ix}^a}&=&
\int\dk{k}\de^{ba}\frac{k_i}{\vec{k}^2}e^{-\imath k\cdot(y-x)},\nonumber\\
\ev{\imath \ro_y^b\imath K_{ix}^a}&=&\ev{\imath \ka_y^b\imath K_{ix}^a}
=\ev{\imath \xi_{\la y}^b\imath K_{ix}^a}=0.
\eea
We see that, as expected, the propagators involving only the vector fields 
are transverse and the only other contributions relate to the Lagrange 
multiplier fields and are purely kinematical in nature.  There is one 
subtlety to the above and that is that whilst the equations for 
$\ev{\imath\xi_{\la}\imath J}$ and $\ev{\imath\xi_{\ta}\imath K}$ are exact 
in the presence of sources, all other equations refer implicitly to the case 
where the sources are set to zero and the discrete parity symmetry has been 
applied (see later for a more complete discussion).

In the same fashion, let us consider the $\la$-equation, \eq{eq:ladse}, (and 
similarly the $\ta$-equation, \eq{eq:tadse}) in terms of proper Green's 
functions.  This equation does contain the same information as its 
counterpart for connected Green's functions, but clearly has a different 
character.  The only non-zero functional derivative is
\be
\ev{\imath A_{iy}^b\imath\la_x^a}
=\int\dk{k}\de^{ba}k_ie^{-\imath k\cdot(y-x)},
\ee
all others vanishing, \emph{even in the presence of sources}.  This applies 
to all proper $n$-point functions involving the $\la$-field.  Similarly, we 
have the only non-vanishing proper function involving the $\ta$-field:
\be
\ev{\imath \pi_{iy}^b\imath\ta_x^a}
=\int\dk{k}\de^{ba}k_ie^{-\imath k\cdot(y-x)}.
\ee
That there are no proper $n$-point functions involving functional 
derivatives with respect to the Lagrange multiplier fields apart from the 
two special cases above leads to an important facet concerning the \DS 
equations -- there will be no self-energy terms involving derivatives with 
respect to the $\la$- or $\ta$-fields since they have no proper vertices, 
despite the fact that the propagators associated with these fields may be 
non-trivial.

Next, let us turn to the $\phi$-based equation of motion in terms of 
connected Green's functions, \eq{eq:phidse}.  There are only two 
non-vanishing functional derivatives and we can write down the solutions as 
before
\bea
\ev{\imath K_{iy}^b\imath\xi_{\ta x}^a}&=&
\int\dk{k}\de^{ba}\frac{(-k_i)}{\vec{k}^2}e^{-\imath k\cdot(y-x)}
,\nonumber\\
\ev{\imath\ka_y^b\imath\xi_{\ta x}^a}&=&
\int\dk{k}\de^{ba}\frac{\imath}{\vec{k}^2}e^{-\imath k\cdot(y-x)}
,\nonumber\\
\ev{\imath J_{iy}^b\imath\xi_{\ta x}^a}&=&
\ev{\imath\ro_y^b\imath\xi_{\ta x}^a}=
\ev{\imath\xi_{\la x}^b\imath\xi_{\ta x}^a}=
\ev{\imath\xi_{\ta x}^b\imath\xi_{\ta x}^a}=0.
\eea
Notice that \emph{all} of the connected Green's functions involving the 
$\ta$-field are now known and are purely kinematical in nature.  In terms of 
proper Green's functions, we consider the $\phi$-based equation of motion, 
\eq{eq:phidse0}.  Recognising that functional derivatives with respect to 
the $\la$- and $\ta$-fields yield no more information, we can omit them from 
the current discussion.  The equation tells us that given a proper Green's 
function involving $\pi$, we can immediately construct the corresponding 
functional derivative with respect to $\phi$.  We can thus conclude that as 
far as the proper Green's functions are concerned, derivatives with repect 
to the $\phi$-field are redundant.

Finally, let us consider the equation derived from the BRS transform in 
terms of connected Green's functions, \eq{eq:stid0}.  Since the 
ghost/antighost fields must come in pairs, we may take the functional 
derivative of this with respect to $\imath\et_z^d$ and subsequently set the 
ghost sources to zero whilst considering only the rest.  We get:
\bea
\lefteqn{\frac{\imath}{g}\ev{\imath\xi_{\la z}^d}=
\int\dx{x}\left\{
\frac{1}{g}(\pd_x^0\ro_x^a)\ev{\imath\ov{\et}_x^a\imath\et_z^d}
-f^{acb}\ro_x^a\left[\ev{\imath\ro_x^c\imath\ov{\et}_x^b\imath\et_z^d}
+\ev{\imath\ro_x^c}\ev{\imath\ov{\et}_x^b\imath\et_z^d}\right]\right.}
\nonumber\\&&
-\frac{1}{g}(\nabla_{ix}J_{ix}^a)\ev{\imath\ov{\et}_x^a\imath\et_z^d}
-f^{acb}J_{ix}^a\left[\ev{\imath J_{ix}^c\imath\ov{\et}_x^b\imath\et_z^d}
+\ev{\imath J_{ix}^c}\ev{\imath\ov{\et}_x^b\imath\et_z^d}\right]
\nonumber\\&&
+f^{abc}\left[K_{ix}^a-\frac{\nabla_{ix}}{(-\nabla_x^2)}
(\ka_x^a-\nabla_{jx}K_{jx}^a)\right]\times
\nonumber\\&&\left.
\left[\ev{\imath K_{ix}^c\imath\ov{\et}_x^b\imath\et_z^d}
-\int\dx{y}\de(x-y)\nabla_{ix}\ev{\imath\ka_x^c\imath\ov{\et}_y^b
\imath\et_z^d}+\ev{\imath K_{ix}^c}\ev{\imath\ov{\et}_x^b\imath\et_z^d}
-\ev{\imath\ov{\et}_x^b\imath\et_z^d}\nabla_{ix}\ev{\imath\ka_x^c}\right]
\right\}.
\label{eq:stid1}
\eea
For now, the pertinent information from this identity comes from taking the 
functional derivative with respect to the source $\xi_{\la}$ and setting all 
sources to zero.  The result is
\be
\ev{\imath\xi_{\la\w}^e\imath\xi_{\la z}^d}=0.
\ee
We notice that all other functional derivatives lead to non-trivial 
relations involving interaction terms.  This includes both the 
$\ev{\imath\ro\imath\xi_\la}$ and the $\ev{\imath\ka\imath\xi_\la}$ 
connected Green's functions and we conclude that these functions are not 
merely kinematical factors as one might expect from quantities involving 
Lagrange multiplier fields.  We shall return to this topic at a later stage.

\section{Feynman (and Other) Rules}
\setcounter{equation}{0}
Whilst it is entirely possible to deduce the complete set of Feynman rules 
directly from the action, we shall follow a slightly less obvious path 
here.  We derive not only the basic Feynman rules but collect all the 
tree-level (and incidently primitively divergent) quantities that will be of 
interest.  This means that in addition to the tree-level propagators (i.e., 
connected two-point Green's functions) and proper vertices (i.e., proper 
three- and four-point functions) we derive also the proper two-point 
functions.  The reason for this is that (as will be discussed in some detail 
later) the connected and proper two-point functions are not related in the 
usual way as inverses of one another.  The tree-level quantities of interest 
can be easily derived from the respective equations of motion.  Indeed, 
recalling the previous section, some are already known exactly and we will 
not need to discuss them further.

Before beginning, let us highlight a basic feature of the Fourier transform 
to momentum space.  We know the commutation or anti-commutation rules for 
our fields/sources and this will lead to the simplification that we need only 
consider combinations of fields/sources and let the commutation rules take 
care of the permutations.  However, momentum assignments must be uniformly 
applied and this leads to some non-trivial relations.  Consider firstly the 
generic proper two-point function $\ev{\imath\Phi_\al(x)\imath\Phi_\ba(y)}$ 
where we have $\Phi_\ba(y)\Phi_\al(x)=\et\Phi_\al(x)\Phi_\ba(y)$ with 
$\et=\pm1$.  We then have
\be
\ev{\imath\Phi_\al(x)\imath\Phi_\ba(y)}
=\et\ev{\imath\Phi_\ba(y)\imath\Phi_\al(x)}
\ee
such that in momentum space
\be
\G_{\al\ba}(k)=\et\G_{\ba\al}(-k).
\ee
A similar argument applies for connected two-point functions.  The situation 
for proper three-point functions is slightly less complicated since all 
momenta are defined as incoming.  Indeed, we have (the $\de$-function 
expressing momentum conservation comes about because of translational 
invariance)
\be
\ev{\imath\Phi_\al\imath\Phi_\ba\imath\Phi_\ga}
=\int\dk{k_\al}\dk{k_\ba}\dk{k_\ga}(2\pi)^4\de(k_\al+k_\ba+k_\ga)
\G_{\al\ba\ga}(k_\al,k_\ba,k_\ga)
e^{-\imath k_\al\cdot x_\al-\imath k_\ba\cdot x_\ba-\imath k_\ga\cdot x_\ga}
\ee
such that, for example
\be
\G_{\ba\al\ga}(k_\ba,k_\al,k_\ga)
=\et_{\al\ba}\G_{\al\ba\ga}(k_\al,k_\ba,k_\ga)
\ee
where $\et_{\al\ba}$ refers to the sign incurred when swapping $\al$ and 
$\ba$.

Let us now consider the connected two-point functions.  Setting the coupling 
to zero in the equations of motion (listed in Appendix~\ref{app:eom}) that 
involve interaction terms gives us the following non-trivial relations (the 
superscript $\ev{}^{(0)}$ denotes the tree-level quantity)
\bea
\ro_x^a-\xi_{\ta x}^a&=&-\nabla_x^2\ev{\imath\ka_x^a}^{(0)},\nonumber\\
J_{ix}^a&=&-\nabla_{ix}\ev{\imath\xi_{\la x}^a}^{(0)}
+\pd_x^0\ev{\imath K_{ix}^a}^{(0)}
-\pd_x^0\nabla_{ix}\ev{\imath\ka_x^a}^{(0)}
-\left[\de_{ij}\nabla_x^2+\nabla_{ix}\nabla_{jx}\right]
\ev{\imath J_{jx}^a}^{(0)},\nonumber\\
\et_x^a&=&\nabla_x^2\ev{\imath\ov{\et}_x^a}^{(0)},\nonumber\\
K_{ix}^a&=&-\nabla_{ix}\ev{\imath\xi_{\ta x}^a}^{(0)}
+\ev{\imath K_{ix}^a}^{(0)}-\nabla_{ix}\ev{\imath\ka_x^a}^{(0)}
-\pd_x^0\ev{\imath J_{ix}^a}^{(0)}-\nabla_{ix}\ev{\imath\ro_x^a}^{(0)}.
\eea
Clearly, the ghost propagator is distinct from the rest, since the ghost 
field must appear with its antighost counterpart.  The treel-level ghost 
propagator is
\be
W_{\ov{c}c}^{(0)ab}(k)=-\de^{ab}\frac{\imath}{\vec{k}^2}.
\ee
The remaining tree-level propagators in momentum space (without the common 
color factor $\de^{ab}$) are summarised in Table~\ref{tab:w0}.  Those 
entries that are underlined are the exact relations considered previously.

\begin{table}
\begin{tabular}{|c||c|c||c|c||c|c|}\hline
$W$&$A_j$&$\pi_j$&$\si$&$\phi$&$\la$&$\ta$\\\hline\rule[-2.4ex]{0ex}{5.5ex}
$A_i$&$t_{ij}(k)\frac{i}{(k_0^2-\vec{k}^2)}$&
$t_{ij}(k)\frac{(-k^0)}{(k_0^2-\vec{k}^2)}$&$\underline{0}$&$\underline{0}$&
$\underline{\frac{(-k_i)}{\vec{k}^2}}$&$\underline{0}$\\
\hline\rule[-2.4ex]{0ex}{5.5ex}
$\pi_i$&$t_{ij}(k)\frac{k^0}{(k_0^2-\vec{k}^2)}$&
$t_{ij}(k)\frac{\imath\vec{k}^2}{(k_0^2-\vec{k}^2)}$&$\underline{0}$&
$\underline{0}$&$\underline{0}$&$\underline{\frac{(-k_i)}{\vec{k}^2}}$\\
\hline\hline\rule[-2.4ex]{0ex}{5.5ex}
$\si$&$\underline{0}$&$\underline{0}$&$\frac{\imath}{\vec{k}^2}$&
$\frac{(-\imath)}{\vec{k}^2}$&$\frac{(-k^0)}{\vec{k}^2}$&$\underline{0}$\\
\hline\rule[-2.4ex]{0ex}{5.5ex}
$\phi$&$\underline{0}$&$\underline{0}$&$\frac{(-\imath)}{\vec{k}^2}$&0&0&
$\underline{\frac{\imath}{\vec{k}^2}}$\\
\hline\hline\rule[-2.4ex]{0ex}{5.5ex}
$\la$&$\underline{\frac{k_j}{\vec{k}^2}}$&$\underline{0}$&
$\frac{k^0}{\vec{k}^2}$&0&$\underline{0}$&$\underline{0}$\\
\hline\rule[-2.4ex]{0ex}{5.5ex}
$\ta$&$\underline{0}$&$\underline{\frac{k_j}{\vec{k}^2}}$&$\underline{0}$&
$\underline{\frac{\imath}{\vec{k}^2}}$&$\underline{0}$&$\underline{0}$\\
\hline
\end{tabular}
\caption{\label{tab:w0}Tree-level propagators (without color factors) in 
momentum space.  Underlined entries denote exact results.}
\end{table}

We can repeat the analysis for the tree-level proper two-point functions.  
The relevant equations are:
\bea
\ev{\imath\si_x^a}^{(0)}-\ev{\imath\ta_x^a}^{(0)}&=&
\nabla_x^2\phi_x^a,\nonumber\\
\ev{\imath A_{ix}^a}^{(0)}&=&\nabla_{ix}\la_x^a-\pd_x^0\pi_{ix}^a
+\pd_x^0\nabla_{ix}\phi_x^a+\left[\de_{ij}\nabla_x^2
-\nabla_{ix}\nabla_{jx}\right]A_{jx}^a,\nonumber\\
\ev{\imath\ov{c}_x^a}^{(0)}&=&-\nabla_x^2c_x^a,\nonumber\\
\ev{\imath\pi_{ix}^a}^{(0)}&=&\nabla_{ix}\ta_x^a-\pi_{ix}^a
+\nabla_{ix}\phi_x^a+\pd_x^0A_{ix}^a+\nabla_{ix}\si_x^a.
\eea
The ghost proper two-point function is
\be
\G_{\ov{c}c}^{(0)ab}(k)=\de^{ab}\imath\vec{k}^2.
\ee
The remaining proper two-point functions are summarised in 
Table~\ref{tab:g0} where the reader is reminded that all proper functions 
involving derivatives with respect to the $\phi$-field can be constructed 
from the corresponding $\pi$ derivative.

\begin{table}
\begin{tabular}{|c||c|c||c|c||c|c|}\hline
$\G$&$A_j$&$\pi_j$&$\si$&$\phi$&$\la$&$\ta$\\
\hline\rule[-2.4ex]{0ex}{5.5ex}
$A_i$&$t_{ij}(k)\imath\vec{k}^2$&$\de_{ij}k^0$&0&
$\left\{-\imath k^0k_i\right\}$&$\underline{k_i}$&$\underline{0}$\\
\hline\rule[-2.4ex]{0ex}{5.5ex}
$\pi_i$&$-k^0\de_{ij}$&$\imath\de_{ij}$&$k_i$&$\left\{k_i\right\}$&
$\underline{0}$&$\underline{k_i}$\\
\hline\hline\rule[-2.4ex]{0ex}{5.5ex}
$\si$&0&$-k_j$&0&$\left\{\imath\vec{k}^2\right\}$&$\underline{0}$&
$\underline{0}$\\
\hline\rule[-2.4ex]{0ex}{5.5ex}
$\phi$&$\left\{-\imath k^0k_j\right\}$&$\left\{-k_j\right\}$&
$\left\{\imath\vec{k}^2\right\}$&$\left\{\imath\vec{k}^2\right\}$&
$\underline{0}$&$\underline{0}$\\
\hline\hline\rule[-2.4ex]{0ex}{5.5ex}
$\la$&$\underline{-k_j}$&$\underline{0}$&$\underline{0}$&$\underline{0}$&
$\underline{0}$&$\underline{0}$\\
\hline\rule[-2.4ex]{0ex}{5.5ex}
$\ta$&$\underline{0}$&$\underline{-k_j}$&$\underline{0}$&$\underline{0}$&
$\underline{0}$&$\underline{0}$\\\hline
\end{tabular}
\caption{\label{tab:g0}Tree-level proper two-point functions (without color 
factors) in momentum space.  Underlined entries denote exact results.  
Bracketed quantities refer to functions that are fully determined by others.}
\end{table}

Determining the tree-level vertices (three- and four-point proper Green's 
functions) follows the same pattern as for the two-point functions.  They 
follow by isolating the parts of the equations of motion that have explicit 
factors of the coupling $g$ and functionally differentiating.  In momentum 
space (defining all momenta to be incoming), we have
\bea
\G_{\pi\si A ij}^{(0)abc}&=&-gf^{abc}\de_{ij},\nonumber\\
\G_{3A ijk}^{(0)abc}(p_a,p_b,p_c)&=&
-\imath gf^{abc}\left[\de_{ij}(p_a-p_b)_k+\de_{jk}(p_b-p_c)_i
+\de_{ki}(p_c-p_a)_j\right],\nonumber\\
\G_{4A ijkl}^{(0)abcd}&=&-\imath g^2\left\{\de_{ij}\de_{kl}
\left[f^{ace}f^{bde}-f^{ade}f^{cbe}\right]
+\de_{ik}\de_{jl}\left[f^{abe}f^{cde}-f^{ade}f^{bce}\right]
+\de_{il}\de_{jk}\left[f^{ace}f^{dbe}f^{abe}f^{cde}\right]\right\}
,\nonumber\\
\G_{\ov{c}cA i}^{(0)abc}(p_{\ov{c}},p_c,p_A)&=&-\imath gf^{abc}p_{\ov{c}i}.
\eea
We notice that all the tree-level vertices are independent of the energy.  
In addition, there is a tree-level vertex involving $\phi$ that can be 
constructed from its counterpart involving $\vec{\pi}$ and that reads:
\be
\G_{\phi\si Ai}^{(0)abc}(p_\phi,p_\si,p_A)
=\imath p_{\phi j}\G_{\pi\si Aji}^{(0)abc}=-\imath gf^{abc}p_{\phi i}.
\ee
This vertex has exactly the same form as the ghost-gluon vertex with the 
incoming $\phi$-momentum playing the same role as the incoming 
$\ov{c}$-momentum.  It is worth mentioning that the ghost-gluon, three- and 
four-gluon vertices are identical to the Landau gauge forms except that only 
the spatial components of the vectors are present.

Let us now discuss the cancellation of the ghost (energy divergent) sector.  
In any Feynman diagram containing a closed ghost loop, there will be an 
associated energy divergence.  It is a general result that associated with 
any closed loop involving Grassmann-valued fields (ghosts or fermions) there 
will be a factor of $(-1)$.  However, Green's functions are given by the sum 
of all possible contributing Feynman diagrams.  Since the Feynman rules for 
$W_{\si\phi}$ and $\G_{\phi\si A}$ are identical to $W_{\ov{c}c}$ and 
$\G_{\ov{c}cA}$ we will have, for each closed ghost loop, another loop 
involving scalar fields without the factor $(-1)$.  Even before performing 
the loop integration (and regularisation) the integrands of the two diagrams 
will cancel exactly.  In this way we see that the energy divergences coming 
from the ghost sector will be eliminated, as expected given that the 
Faddeev-Popov determinant can formally be cancelled.  There is one caveat to 
this.  Whilst we have shown that the energy divergences coming from the 
ghost sector have been eliminated, we have not shown that the remaining 
loops involving scalar fields are free of energy divergences (although a 
quick glance at the form of the \DS equations later will suffice to see that 
this is the case at leading order).  We propose to look further into this in 
a future publication.

\section{Decomposition of Two-Point Functions}
\setcounter{equation}{0}
In order to constrain the possible form of the two-point functions under 
investigation we can utilise information about discrete symmetries.  We 
consider time-reversal and parity and we know that Yang-Mills theory 
respects both.  Under time-reversal the generic field 
$\Phi_\al(x^0,\vec{x})$ is transformed as follows:
\be
\Phi_\al(x^0,\vec{x})=\et_\al\Phi_\al(-x^0,\vec{x})
\ee
where $\et_\al=\pm1$.  Since the action, \eq{eq:act}, is invariant under 
time-reversal (it is a pure number) then by considering each term in turn, 
we deduce that
\be
\et_A=\et_\la=1,\;\;\;\;\et_\pi=\et_\ta=\et_\phi=\et_\si=-1,\;\;\;\;
\et_{\ov{c}}=\et_c=\pm1.
\ee
The sources have the same transformation properties as the field.  These 
properties allow us to extract information about the energy dependence of 
Green's functions.  For instance, we have that
\be
\G_{A\pi ij}^{ab}(k^0,\vec{k})=-\G_{A\pi ij}^{ab}(-k^0,\vec{k}),
\ee
from which one can infer that
\be
\G_{A\pi ij}^{ab}(k^0,\vec{k})=\de^{ab}k^0\G_{A\pi ij}(k_0^2,\vec{k})
\ee
(the sign convention is chosen to match the perturbative results).  Aside 
from $\G_{A\phi}$ which is unambiguously related to $\G_{A\pi}$, the only 
other proper two-point function that carries the external factor $k^0$ is
\be
\G_{A\si i}^{ab}(k^0,\vec{k})=\de^{ab}k^0\G_{A\si i}(k_0^2,\vec{k}).
\ee
Having extracted the explicit factors of $k^0$ in the proper two-point 
funcitons, the (as yet) unknown functions that multiply them are functions 
of $k_0^2$.  Turning to the propagators, we assign the factor $-k^0$ to 
$W_{A\pi}$, $W_{\si\la}$ and $W_{\phi\la}$.

The second discrete symmetry of interest is parity whereby
\be
\Phi_\al(x^0,\vec{x})=\et_\al\Phi_\al(x^0,-\vec{x})
\ee
where again, $\et_\al=\pm1$.  Again the action is invariant and we deduce 
that
\be
\et_A=\et_\pi=-1,\;\;\;\;\et_\si=\et_\phi=\et_\la=\et_\ta=1,\;\;\;\;
\et_{\ov{c}}=\et_c=\pm1
\ee
with the sources transforming as the fields.  This symmetry is rather more 
obvious than time-reversal.  The physical sense is that for every vector 
field (and with an associated spatial index) we have some explicit vector 
factor (again with the associated spatial index).  Where the vector fields 
$\vec{A}$ and $\vec{\pi}$ occur in the propagators, we use the 
transversality conditions from earlier to see that the vector-scalar 
propagators must vanish, except those involving the apropriate Lagrange 
multiplier field.

What the above tells us is how to construct the most general allowed forms 
of the two-point functions.  The dressing functions are scalar functions of 
the positive, scalar arguments $k_0^2$ and $\vec{k}^2$.  We summarise the 
results in Tables~\ref{tab:w1} and \ref{tab:g1}.  The ghost propagator is 
written $W_{\ov{c}c}^{ab}(k)=-\de^{ab}\imath D_c/\vec{k}^2$.  We have nine 
unknown propagator dressing functions.  Inlcuding the proper ghost two-point 
function $\G_{\ov{c}c}^{ab}(k)=\de^{ab}\imath\vec{k^2}\G_c$ we see that 
there are ten proper two-point dressing functions.  The extra functions 
comes about because we have used only the propagator form of the identity 
\eq{eq:stid1} to eleminate $W_{\la\la}$.

\begin{table}
\begin{tabular}{|c||c|c||c|c||c|c|}\hline
$W$&$A_j$&$\pi_j$&$\si$&$\phi$&$\la$&$\ta$\\
\hline\rule[-2.4ex]{0ex}{5.5ex}
$A_i$&$t_{ij}(k)\frac{\imath D_{AA}}{(k_0^2-\vec{k}^2)}$&
$t_{ij}(k)\frac{(-k^0)D_{A\pi}}{(k_0^2-\vec{k}^2)}$&0&0&
$\frac{(-k_i)}{\vec{k}^2}$&0\\
\hline\rule[-2.4ex]{0ex}{5.5ex}
$\pi_i$&$t_{ij}(k)\frac{k^0D_{A\pi}}{(k_0^2-\vec{k}^2)}$&
$t_{ij}(k)\frac{\imath\vec{k}^2D_{\pi\pi}}{(k_0^2-\vec{k}^2)}$&0&0&0&
$\frac{(-k_i)}{\vec{k}^2}$\\
\hline\hline\rule[-2.4ex]{0ex}{5.5ex}
$\si$&0&0&$\frac{\imath D_{\si\si}}{\vec{k}^2}$&
$\frac{-\imath D_{\si\phi}}{\vec{k}^2}$&
$\frac{(-k^0)D_{\si\la}}{\vec{k}^2}$&0\\
\hline\rule[-2.4ex]{0ex}{5.5ex}
$\phi$&0&0&$\frac{-\imath D_{\si\phi}}{\vec{k}^2}$&
$\frac{-\imath D_{\phi\phi}}{\vec{k}^2}$&
$\frac{(-k^0)D_{\phi\la}}{\vec{k}^2}$&$\frac{\imath}{\vec{k}^2}$\\
\hline\hline\rule[-2.4ex]{0ex}{5.5ex}
$\la$&$\frac{k_j}{\vec{k}^2}$&0&$\frac{k^0D_{\si\la}}{\vec{k}^2}$&
$\frac{k^0D_{\phi\la}}{\vec{k}^2}$&0&0\\
\hline\rule[-2.4ex]{0ex}{5.5ex}
$\ta$&0&$\frac{k_j}{\vec{k}^2}$&0&$\frac{\imath}{\vec{k}^2}$&0&0\\
\hline
\end{tabular}
\caption{\label{tab:w1}General form of propagators in momentum space.  The 
global color factor $\de^{ab}$ has been extracted.  All unknown functions 
$D_{\al\ba}$ are dimensionless, scalar functions of $k_0^2$ and $\vec{k}^2$.}
\end{table}

\begin{table}
\begin{tabular}{|c||c|c||c|c||c|c|}\hline
$\G$&$A_j$&$\pi_j$&$\si$&$\phi$&$\la$&$\ta$\\
\hline\rule[-2.4ex]{0ex}{5.5ex}
$A_i$&$t_{ij}(\vec{k})\imath\vec{k}^2\G_{AA}+\imath k_ik_j\ov{\G}_{AA}$&
$k^0\left(\de_{ij}\G_{A\pi}+l_{ij}(\vec{k})\ov{\G}_{A\pi}\right)$&
$-\imath k^0k_i\G_{A\si}$&$\left\{-\imath k^0k_i\left(\G_{A\pi}
+\ov{\G}_{A\pi}\right)\right\}$&$k_i$&0\\
\hline\rule[-2.4ex]{0ex}{5.5ex}
$\pi_i$&$-k^0\left(\de_{ij}\G_{A\pi}+l_{ij}(\vec{k})\ov{\G}_{A\pi}\right)$&
$\imath\de_{ij}\G_{\pi\pi}+\imath l_{ij}(\vec{k})\ov{\G}_{\pi\pi}$&
$k_i\G_{\pi\si}$&
$\left\{k_i\left(\G_{\pi\pi}+\ov{\G}_{\pi\pi}\right)\right\}$&0&$k_i$\\
\hline\hline\rule[-2.4ex]{0ex}{5.5ex}
$\si$&$-\imath k^0k_j\G_{A\si}$&$-k_j\G_{\pi\si}$&
$\imath\vec{k}^2\G_{\si\si}$&$\left\{\imath\vec{k}^2\G_{\pi\si}\right\}$&0&
0\\\hline\rule[-2.4ex]{0ex}{5.5ex}
$\phi$&$\left\{-\imath k^0k_j\left(\G_{A\pi}+\ov{\G}_{A\pi}\right)\right\}$&
$\left\{-k_j\left(\G_{\pi\pi}+\ov{\G}_{\pi\pi}\right)\right\}$&
$\left\{\imath\vec{k}^2\G_{\pi\si}\right\}$&
$\left\{-\imath\vec{k}^2\left(\G_{\pi\pi}+\ov{\G}_{\pi\pi}\right)
\right\}$&0&0\\\hline\hline\rule[-2.4ex]{0ex}{5.5ex}
$\la$&$-k_j$&0&0&0&0&0\\\hline\rule[-2.4ex]{0ex}{5.5ex}
$\ta$&0&$-k_j$&0&0&0&0\\\hline
\end{tabular}
\caption{\label{tab:g1}General form of the proper two-point functions in 
momentum space.   The global color factor $\de^{ab}$ has been extracted.  
All unknown functions $D_{\al\ba}$ are dimensionless, scalar functions of 
$k_0^2$ and $\vec{k}^2$. Bracketed quantities refer to functions that are 
fully determined by others.}
\end{table}

Obviously the propagator and proper two-point dressing functions are related 
via the Legendre transform.  Whereas in covariant gauges this relationship 
is merely an inversion, in our case there is considerably more detail.  The 
connection between the connected and proper two-point functions stems from 
the observation that
\be
\frac{\de\imath J_\ba}{\de\imath J_\al}=\de_{\al\ba}
=-\imath\frac{\de}{\de\imath J_\al}\ev{\imath\Phi_\ba}
=\frac{\de\Phi_\ga}{\de\imath J_\al}\ev{\imath\Phi_\ga\imath\Phi_\ba}
=\ev{\imath J_\al\imath J_\ga}\ev{\imath\Phi_\ga\imath\Phi_\ba}.
\label{eq:leg}
\ee
(Recall here that there is an implicit summation over all discrete indices 
and integration over continuous variables labelled by $\ga$.)  The ghost 
two-point functions are somewhat special in that once sources are set to 
zero, only ghost-antighost pairs need be considered.  The above relation 
becomes
\be
\int d^4z\ev{\imath\ov{\et}_x^a\imath\et_z^c}
\ev{\imath\ov{c}_z^c\imath c_y^b}=\de^{ab}\de(x-y).
\label{eq:ghleg}
\ee
Fourier transforming to momentum space and using the decomposition from 
above, we get that
\be
D_c(k_0^2,\vec{k}^2)\G_c(k_0^2,\vec{k}^2)=1
\ee
showing that the ghost propagator dressing function is simply the inverse of 
the ghost proper two-point function.  Turning to the rest, we are faced with 
a problem akin to matrix inversion in order to see the connection since the 
sum over all the different possible sources/fields labelled by $\ga$ is 
non-trivial in the above general formula.  The decompositions of the 
two-point function do however mitigate the complexity somewhat.  We tabulate 
the possible combinations of terms in Table~\ref{tab:leg0}.

\begin{table}[t]
\begin{tabular}{|c||c|c||c|c||c|c|}\hline
$\al$,$\ba$&$\vec{A}$&$\vec{\pi}$&$\si$&$\phi$&$\la$&$\ta$\\
\hline\hline
$\vec{A}$&$\vec{A}$,$\vec{\pi}$,$\la$&$\vec{A}$,$\vec{\pi}$&
$\vec{A}$,$\vec{\pi}$&$\vec{A}$,$\vec{\pi}$&$\vec{A}$&$\vec{\pi}$\\
\hline
$\vec{\pi}$&$\vec{A}$,$\vec{\pi}$&$\vec{A}$,$\vec{\pi}$,$\ta$&
$\vec{A}$,$\vec{\pi}$&$\vec{A}$,$\vec{\pi}$&$\vec{A}$&$\vec{\pi}$\\
\hline\hline
$\si$&$\si$,$\phi$,$\la$&$\si$,$\phi$&$\si$,$\phi$&$\si$,$\phi$&---&---\\
\hline
$\phi$&$\si$,$\phi$,$\la$&$\si$,$\phi$,$\ta$&$\si$,$\phi$&$\si$,$\phi$&
---&---\\\hline\hline
$\la$&$\vec{A}$,$\si$,$\phi$&$\vec{A}$,$\si$,$\phi$&$\vec{A}$,$\si$,$\phi$&
$\vec{A}$,$\si$,$\phi$&$\vec{A}$&---\\\hline
$\ta$&$\vec{\pi}$,$\phi$&$\vec{\pi}$,$\phi$&$\vec{\pi}$,$\phi$&
$\vec{\pi}$,$\phi$&---&$\vec{\pi}$\\\hline
\end{tabular}
\caption{\label{tab:leg0}Possible terms for the equations relating 
propagator and proper two-point functions stemming from the Legendre 
transform.  Entries denote the allowed field types $\ga$ in \eq{eq:leg}.}
\end{table}

We start by considering the top left components of Table~\ref{tab:leg0} 
involving only $\vec{A}$, $\vec{\pi}$ and the known functions with $\la$ and 
$\ta$.  After decomposition, we have (suppressing the common argument $k$)
\bea
k_0^2D_{A\pi}\G_{A\pi}-\vec{k}^2D_{AA}\G_{AA}&=&k_0^2-\vec{k}^2,\nonumber\\
k_0^2D_{A\pi}\G_{A\pi}-\vec{k}^2D_{\pi\pi}\G_{\pi\pi}&=&k_0^2-\vec{k}^2
,\nonumber\\
D_{AA}\G_{A\pi}-D_{A\pi}\G_{\pi\pi}&=&0,\nonumber\\
D_{A\pi}\G_{AA}-D_{\pi\pi}\G_{A\pi}&=&0.
\eea
We can thus express the propagator functions $D$ in terms of the proper 
two-point functions $\G$ and we have
\bea
D_{AA}&=&\frac{\left(k_0^2-\vec{k}^2\right)\G_{\pi\pi}}{
\left(k_0^2\G_{A\pi}^2-\vec{k}^2\G_{AA}\G_{\pi\pi}\right)},\nonumber\\
D_{\pi\pi}&=&\frac{\left(k_0^2-\vec{k}^2\right)\G_{AA}}{
\left(k_0^2\G_{A\pi}^2-\vec{k}^2\G_{AA}\G_{\pi\pi}\right)},\nonumber\\
D_{A\pi}&=&\frac{\left(k_0^2-\vec{k}^2\right)\G_{A\pi}}{
\left(k_0^2\G_{A\pi}^2-\vec{k}^2\G_{AA}\G_{\pi\pi}\right)}.
\eea
Clearly, these expressions can be inverted to give the functions $\G$ in 
terms of the functions $D$.  Next, let us consider the central components of 
Table~\ref{tab:leg0} involving only $\si$ and $\phi$.  We get the following 
equations:
\bea
-D_{\si\si}\G_{\si\si}+D_{\si\phi}\G_{\pi\si}&=&1,\nonumber\\
D_{\si\phi}\G_{\pi\si}+D_{\phi\phi}\left(\G_{\pi\pi}
+\ov{\G}_{\pi\pi}\right)&=&1,\nonumber\\
-D_{\si\si}\G_{\pi\si}+D_{\si\phi}\left(\G_{\pi\pi}
+\ov{\G}_{\pi\pi}\right)&=&0,\nonumber\\
D_{\si\phi}\G_{\si\si}+D_{\phi\phi}\G_{\pi\si}&=&0.
\eea
The propagator functions in terms of the proper two-point functions are then
\bea
D_{\si\si}&=&\frac{\left(\G_{\pi\pi}+\ov{\G}_{\pi\pi}\right)}{
\G_{\pi\si}^2-\G_{\si\si}\left(\G_{\pi\pi}+\ov{\G}_{\pi\pi}\right)}
,\nonumber\\
D_{\phi\phi}&=&-\frac{\G_{\si\si}}{\G_{\pi\si}^2-\G_{\si\si}
\left(\G_{\pi\pi}+\ov{\G}_{\pi\pi}\right)},\nonumber\\
D_{\si\phi}&=&\frac{\G_{\pi\si}}{\G_{\pi\si}^2-\G_{\si\si}\left(\G_{\pi\pi}
+\ov{\G}_{\pi\pi}\right)}.
\eea
There are three more equations that are of interest.  These are the $\si-A$, 
$\phi-A$ and $\la-A$ entries of Table~\ref{tab:leg0} and they read:
\bea
D_{\si\si}\G_{A\si}-D_{\si\phi}\left(\G_{A\pi}+\ov{\G}_{A\pi}\right)
+D_{\si\la}&=&0,\\
-D_{\si\phi}\G_{A\si}-D_{\phi\phi}\left(\G_{A\pi}+\ov{\G}_{A\pi}\right)
+D_{\phi\la}&=&0,\\
\ov{\G}_{AA}-\frac{k_0^2}{\vec{k}^2}\left[D_{\si\la}\G_{A\si}
+D_{\phi\la}\left(\G_{A\pi}+\ov{\G}_{A\pi}\right)\right]&=&0.
\eea
What these equations tell us is that $D_{\si\la}$ and $D_{\phi\la}$ are 
related to $\ov{\G}_{A\pi}$ and $\G_{A\si}$ with all other coefficients 
being determined.  $\ov{\G}_{AA}$ is then given as a specific combination 
and is the `extra' proper two-point function alluded to earlier.  However, 
these functions will not be of any real concern since $D_{\si\la}$ and 
$D_{\phi\la}$ do not enter any loop diagrams of the \DS equations.  In 
effect, $\ov{\G}_{A\pi}$, $\G_{A\si}$ and $\ov{\G}_{AA}$ form a consistency 
check on the truncation of the \DS equations since we have that
\be
\ov{\G}_{AA}=\frac{k_0^2}{\vec{k}^2}\left[-D_{\si\si}\G_{A\si}^2
+2D_{\si\phi}\G_{A\si}\left(\G_{A\pi}+\ov{\G}_{A\pi}\right)
+D_{\phi\phi}\left(\G_{A\pi}+\ov{\G}_{A\pi}\right)^2\right].
\ee
It is apparent that unlike covariant gauges, the proper two-point function 
for the gluon is not necessarily transverse.

In summary, leaving the problem of the vertices aside, in order to solve the 
two-point \DS equations we need to calculate seven proper two-point 
functions:
\be
\G_c,\;\;\G_{AA},\;\;\G_{A\pi},\;\;\G_{\pi\pi},\;\;\ov{\G}_{\pi\pi},\;\;
\G_{\si\si},\;\;\G_{\si\pi},
\ee
which will give us the required propagator functions:
\be
D_c,\;\;D_{AA},\;\;D_{A\pi},\;\;D_{\pi\pi},\;\;D_{\si\si},\;\;D_{\si\phi}
,\;\;D_{\phi\phi}.
\ee
The three proper two-point functions $\ov{\G}_{AA}$, $\ov{\G}_{A\pi}$ and 
$\G_{A\si}$ give a consistency check on any truncation scheme but do not 
directly contribute further.

\section{Derivation of the Propagator \DS Equations}
\setcounter{equation}{0}
In this section, we present the explicit derivation of the relevant \DS 
equations for proper two-point functions.

\subsection{Ghost Equations}
As will be shown in this subsection, the ghost sector of the theory plays a 
rather special role.  We will begin by deriving the ghost \DS equation (this 
will serve as a template for the derivation of the other \DS equations).  
With this it is possible to point out two particular features of the ghost 
sector: that the ghost-gluon vertex is UV finite and that the energy ($k^0$ 
component) argument of any ghost line is irrelevant, i.e., that any proper 
function involving ghost fields is independent of the ghost energy.

The derivation of the \DS equation for the ghost proper two-point function 
begins with \eq{eq:ghost0}.  Taking the functional derivative with respect 
to $\imath c_w^d$, using the configuration space definition of the 
tree-level ghost-gluon vertex and omitting terms which will eventually 
vanish when sources are set to zero, we have
\be
\ev{\imath c_w^d\imath\ov{c}_x^a}=\imath\de^{da}\nabla_x^2\de(w-x)
+\int\dx{y}\dx{z}\G_{\ov{c}cAi}^{(0)abc}(x,y,z)\frac{\de}{\de\imath c_w^d}
\ev{\imath\ov{\et}_y^b\imath J_{iz}^c}.
\ee
Using partial differentiation we see that
\be
\frac{\de}{\de\imath c_w^d}\ev{\imath\ov{\et}_y^b\imath J_{iz}^c}
=-\imath\int\dx{v}\ev{\imath J_{iz}^c\imath\ov{\et}_y^b\imath\et_v^e}
\ev{\imath\ov{c}_v^e\imath c_w^d}.
\ee
In the above we have again used the fact that when sources are set to zero, 
the only ghost functions that survive are those with pairs of 
ghost-antighost fields.  Since the ghost fields anticommute, we get that
\be
\ev{\imath\ov{c}_x^a\imath c_w^d}=-\imath\de^{ad}\nabla_x^2\de(x-w)
+\imath\int\dx{y}\dx{z}\dx{v}\G_{\ov{c}cAi}^{(0)abc}(x,y,z)
\ev{\imath J_{iz}^c\imath\ov{\et}_y^b\imath\et_v^e}
\ev{\imath\ov{c}_v^e\imath c_w^d}
\ee
Taking the partial derivative of \eq{eq:ghleg} with respect to 
$\imath J_{iz}^c$ we have (notice that when using partial derivatives here, 
we must include all possible contributions which for clarity are included 
explicitly here):
\be
\int\dx{v}\ev{\imath J_{iz}^c\imath\ov{\et}_y^b\imath\et_v^e}
\ev{\imath\ov{c}_v^e\imath c_w^d}=-\imath\int\dx{v}\dx{u}
\ev{\imath\ov{\et}_y^b\imath\et_v^e}\left\{
\ev{\imath J_{iz}^c\imath J_{ju}^f}
\ev{\imath A_{ju}^f\imath\ov{c}_v^e\imath c_w^d}
+\ev{\imath J_{iz}^c\imath K_{ju}^f}
\ev{\imath\pi_{ju}^f\imath\ov{c}_v^e\imath c_w^d}\right\}.
\ee
Our ghost \DS equation in configuration space is thus
\bea
\lefteqn{\ev{\imath\ov{c}_x^a\imath c_w^d}
=-\imath\de^{ad}\nabla_x^2\de(x-w)}\nonumber\\&&
+\int\dx{y}\dx{z}\dx{v}\dx{u}\G_{\ov{c}cAi}^{(0)abc}(x,y,z)
\ev{\imath\ov{\et}_y^b\imath\et_v^e}\left\{
\ev{\imath J_{iz}^c\imath J_{ju}^f}
\ev{\imath A_{ju}^f\imath\ov{c}_v^e\imath c_w^d}
+\ev{\imath J_{iz}^c\imath K_{ju}^f}
\ev{\imath\pi_{ju}^f\imath\ov{c}_v^e\imath c_w^d}\right\}.
\eea
We Fourier transform this result to get the \DS equation for the proper 
two-point ghost function in momentum space:
\bea
\lefteqn{\G_c^{ad}(k)=\de^{ad}\imath\vec{k}^2}\nonumber\\
&&-\int\left(-\dk{\w}\right)\G_{\ov{c}cAi}^{(0)abc}(k,-\w,\w-k)
W_c^{be}(\w)\left\{W_{AAij}^{cf}(k-\w)\G_{\ov{c}cAj}^{edf}(\w,-k,k-\w)
+W_{A\pi ij}^{cf}(k-\w)\G_{\ov{c}c\pi j}^{edf}(\w,-k,k-\w)\right\}
.\nonumber\\
\label{eq:ghdse0}
\eea
With the convention that the self-energy term on the right-hand side has an 
overall minus sign, we identify $\left(-\dk{\w}\right)$ as the loop 
integration measure in momentum space.

With any two-point \DS equation, it is clear that there are two orderings 
for the functional derivatives on the left-hand side.  In the same way, 
there are three orderings for three-point functions and so on.  This means 
that there are $n$ different equations for the $n$-point proper Green's 
functions, although obviously they all have the same solution and must be 
related in some way.  It is therefore instructive to consider also the 
equation generated by the reverse ordering to see if this will have any 
consequence.  In the ghost case, this means repeating the above analysis but 
starting with the second ghost equation of motion, \eq{eq:ghdse2}.  The 
corresponding \DS equation in momentum space is
\bea
\lefteqn{\G_c^{ad}(k)=\de^{ad}\imath\vec{k}^2}\nonumber\\&&
-\int\left(-\dk{\w}\right)\left\{W_{AAij}^{fc}(\w)
\G_{\ov{c}cAj}^{abc}(k,-k-\w,\w)+W_{A\pi ij}^{fc}(\w)
\G_{\ov{c}c\pi j}^{abc}(k,-k-\w,\w)\right\}W_c^{be}(k+\w)
\G_{\ov{c}cAi}^{(0)edf}(k+\w,-k,-\w).\nonumber\\
\label{eq:ghdse1}
\eea
This equation is formally equivalent to \eq{eq:ghdse0} but we notice that 
the ordering of the dressed vertices is different.  (It is useful to check 
that the two equations are the same by taking both vertices to be bare such 
that the equivalence is manifest.)

\begin{figure}[t]
\includegraphics{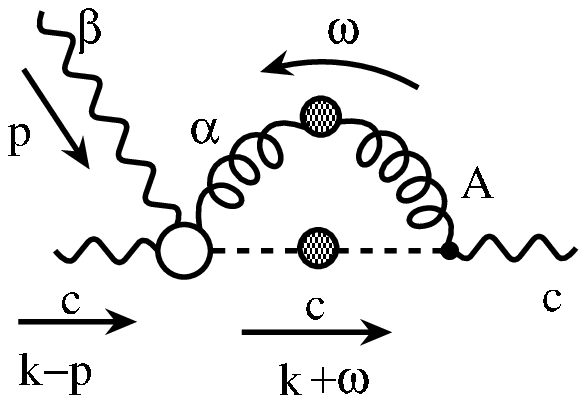}
\caption{\label{fig:ghost} A diagrammatical representation of the 
$\G_{\ov{c}c\ba}(k-p,-k,p)$ proper vertex dressing.  Because of the form of 
the tree-level ghost-gluon vertex and the tranversality of the vector 
propagator, the dressing function vanishes in the limit $k\rightarrow0$.  
Filled blobs denote dressed propagators and empty circles denote dressed 
proper vertex functions.  Wavy lines denote proper functions, springs deonte 
connected (propagator) functions and dashed lines denote the ghost 
propagator.}
\end{figure}

Notice that one of the vertices that form the loop term(s) must be bare.  
This arises naturally through the derivation above and if one considers a 
perturbative expansion it is crucial to avoid overcounting of graphs.  The 
choice of which vertex is bare is arbitrary and related to the fact that 
there are $n$ ways of writing the equation for an $n$-point function.  Given 
that for any loop term we can extract a single bare vertex, for any 
three-point function involving a ghost-antighost pair we will have a loop 
term with the following structure (see also Figure~\ref{fig:ghost}):
\be
\int\dk{\w}\G_{\al\ba\ov{c}cj}^{dgac}(\w,p,k-p,-k-\w)W_c^{ce}(k+\w)
W_{A\al ij}^{fd}(\w)\G_{\ov{c}cAi}^{(0)ebf}(k+\w,-k,-\w).
\ee
Now, since the only propagators involving $A$ are transverse (the $W_{A\la}$ 
propagator is disallowed since no proper vertex function with 
$\la$-derivative exists) the loop term must vanish as $k\rightarrow0$ for 
finite $p$\footnote{The possibility that 
$\G_{\Phi_{\al}\Phi_{\ba}\ov{c}c}^{dgac}(\w,p,k-p,-k-\w)$ is also singular 
in this limit such that the loop remains finite is discounted since at 
finite $\w$ and $p$ this would imply that some colored bound state 
exists.}.  Since the loop term vanishes under some finite, kinematical 
configuration, an UV divergence (which is independent of the kinematical 
configuration) cannot occur and we can can say that this vertex is UV 
finite.  It is tempting to think that such an argument applies to the 
two-point ghost equation, however this is false since whilst the loop term 
vanishes, so does the $\vec{k}^2$ factor that multiplies the rest of the 
equation.

Let us now show that any Green's function involving a ghost-antighost pair 
is independent of the ghost and antighost energies.  The proof of this is 
perturbative in nature.  We notice that both the tree-level ghost propagator 
and the ghost-gluon vertex are independent of the energy.  This means that 
in any one-loop diagram which has at least one internal ghost propagator 
(and hence at least two ghost-gluon vertices) the energy scale associated 
with the ghost propagator is absent.  Using energy conservation another 
energy scale can be eliminated and we choose this to be the antighost 
energy.  At two-loops, we now have the situation whereby the dressed 
internal ghost propagator is again independent of the energy and the dressed 
ghost-gluon vertex only depends on the gluon energy and so the argument can 
be repeated.  This can be applied to all orders in the perturbative 
expansion which completes the proof.  We thus have that in particular
\bea
D_c(k_0^2,\vec{k}^2)&=&D_c(\vec{k}^2)\nonumber\\
\G_{A\ov{c}ci}(k_1,k_2,k_3)&=&\G_{A\ov{c}ci}(k_1,\vec{k}_2,\vec{k}_3).
\eea

\subsection{$\si$-based Equation}
Given the discussion in the previous section about which proper two-point 
functions are relevant, there are only two proper two-point functions 
involving derivatives with respect to $\si$ to consider --- 
$\ev{\imath\si\imath\si}$ and $\ev{\imath\si\imath\pi}$.  Since the 
$\si$-based equation of motion, \eq{eq:sidse0}, involves two interaction terms 
whereas the $\pi$-based equation, \eq{eq:pidse0}, has only one, we use the 
derivatives of \eq{eq:pidse0} to derive the \DS equation for 
$\ev{\imath\si\imath\pi}$ (see next subsection).  We therefore consider the 
functional derivative of \eq{eq:sidse0} with respect to $\imath\si_w^d$, 
after which the sources will be set to zero.  We have, again identifying the 
tree-level vertices,
\be
\ev{\imath\si_w^d\imath\si_x^a}=-\int\dx{y}\dx{z}
\G_{\pi\si Aij}^{(0)cab}(z,x,y)\frac{\de}{\de\imath\si_w^d}
\ev{\imath J_{jy}^b\imath K_{iz}^c}-\int\dx{y}\dx{z}
\G_{\phi\si Ai}^{(0)cab}(z,x,y)\frac{\de}{\de\imath\si_w^d}
\ev{\imath J_{iy}^b\imath\ka_z^c}.
\ee
Using partial differentiation, and with compact notation,
\be
\frac{\de}{\de\imath\si_w^d}\ev{\imath J_{jy}^b\imath K_{iz}^c}
=-\ev{\imath K_{iz}^c\imath J_\al}\ev{\imath J_{jy}^b\imath J_\ba}
\ev{\imath\Phi_\ba\imath\Phi_\al\imath\si_w^d}
\ee
(similarly for the second term).  This gives the \DS equation in 
configuration space:
\bea
\ev{\imath\si_w^d\imath\si_x^a}&=&\int\dx{y}\dx{z}
\G_{\pi\si Aij}^{(0)cab}(z,x,y)\ev{\imath K_{iz}^c\imath J_\al}
\ev{\imath J_{jy}^b\imath J_\ba}
\ev{\imath\Phi_\ba\imath\Phi_\al\imath\si_w^d}\nonumber\\&&
+\int\dx{y}\dx{z}\G_{\phi\si Ai}^{(0)cab}(z,x,y)
\ev{\imath\ka_z^c\imath J_\al}\ev{\imath J_{iy}^b\imath J_\ba}
\ev{\imath\Phi_\ba\imath\Phi_\al\imath\si_w^d}
\eea
Taking the Fourier transform and tidying-up indices, the \DS equation in 
momentum space is thus
\bea
\G_{\si\si}^{ad}(k)&=&-\int\left(-\dk{\w}\right)
\G_{\pi\si Aij}^{(0)cab}(\w-k,k,-\w)W_{A\ba jl}^{be}(\w)
\G_{\ba\al\si lk}^{efd}(\w,k-\w,-k)W_{\al\pi ki}^{fc}(\w-k)\nonumber\\
&&-\int\left(-\dk{\w}\right)\G_{\phi\si Ai}^{(0)cab}(\w-k,k,-\w)
W_{A\ba ij}^{be}(\w)\G_{\ba\al\si j}^{efd}(\w,k-\w,-k)
W_{\al\phi}^{fc}(\w-k).
\label{eq:sidse1}
\eea
A couple of remarks are in order here.  Firstly, there is no bare term on 
the right-hand side because the action, under the first order formalism, is 
linear in $\si$.  Secondly, the implicit summation over the terms labelled 
by $\al$ and $\ba$ means that in fact there are eight possible loop terms 
comprising the self-energy.  However, only two of these involves a 
primitively divergent vertex.  It is an uncomfortable truth that the formal, 
non-local delta function constraint arising from the linearity of the action 
in $\si$ blossoms into a large set of local self-energy integrals.

\subsection{$\pi$-based equations}
Since the $\pi$-based equation of motion, \eq{eq:pidse0}, contains only a 
single interaction term, we favor it to calculate the 
$\ev{\imath A\imath\pi}$ and $\ev{\imath\pi\imath\si}$ proper two-point 
functions (as well as $\ev{\imath\pi\imath\pi}$).  As discussed previously, 
we could in principle calculate these from the $A$-based and $\si$-based 
equations as well in order to check the veracity of any truncations used and 
in fact, this connection may serve useful in elucidating constraints on the 
form of the truncated vertices used.  Perturbatively, all equations will 
provide the same result at any given order.

Using the same techniques as in the last subsection, we get the following 
\DS equations in momentum space:
\bea
\G_{\pi\si i}^{ad}(k)&=&\de^{ad}k_i-\int(-\dk{\w})
\G_{\pi\si Aij}^{(0)abc}(k,-\w,\w-k)W_{\si\ba}^{be}(\w)
\G_{\ba\al\si l}^{efd}(\w,k-\w,-k)W_{\al Alj}^{fc}(\w-k),\\
\G_{\pi Aik}^{ad}(k)&=&-\de^{ad}k^0\de_{ik}-\int(-\dk{\w})
\G_{\pi\si Aij}^{(0)abc}(k,-\w,\w-k)W_{\si\ba}^{be}(\w)
\G_{\ba\al Alk}^{efd}(\w,k-\w,-k)W_{\al Alj}^{fc}(\w-k),\\
\G_{\pi\pi ik}^{ad}(k)&=&\imath\de^{ad}\de_{ik}-\int(-\dk{\w})
\G_{\pi\si Aij}^{(0)abc}(k,-\w,\w-k)W_{\si\ba}^{be}(\w)
\G_{\ba\al\pi lk}^{efd}(\w,k-\w,-k)W_{\al Alj}^{fc}(\w-k).
\eea
Again, notice that the summation over the allowed types of fields indicated 
by $\al$ and $\ba$ leads to multiple possibilities.

\subsection{$A$-based equation}
Using the tree-level forms for the vertices and discarding those terms which 
will eventually vanish when sources are set to zero, it is possible to 
rewrite \eq{eq:adse0} as
\bea
\ev{\imath A_{ix}^a}&=&
\left[\de_{ij}\nabla_x^2-\nabla_{ix}\nabla_{jx}\right]A_{jx}^a
+\int\dx{y}\dx{z}\G_{\ov{c}cAi}^{(0)bca}(z,y,x)
\ev{\imath\ov{\et}_y^c\imath\et_z^b}-\int\dx{y}\dx{z}
\G_{\phi\si Ai}^{(0)bca}(z,y,x)\ev{\imath\ro_y^c\imath\ka_z^b}
\nonumber\\&&
-\int\dx{y}\dx{z}\G_{\pi\si Aji}^{(0)bca}(z,y,x)
\ev{\imath\ro_y^c\imath K_{jz}^b}
-\int\dx{y}\dx{z}\ha\G_{3Akji}^{(0)bca}(z,y,x)
\ev{\imath J_{jy}^c\imath J_{kz}^b}
\nonumber\\&&
-\int\dx{y}\dx{z}\dx{w}\frac{1}{6}\G_{4Alkji}^{(0)dcba}(w,z,y,x)
\left[3\imath A_{jy}^b\ev{\imath J_{kz}^c\imath J_{lw}^d}
+\imath\ev{\imath J_{jy}^b\imath J_{kz}^c\imath J_{lw}^d}\right].
\eea
Functionally differentiating this with respect to $A$ and proceeding as 
before, noting the following for the four-gluon connected vertex
\bea
\imath\frac{\de}{\de\imath A_{mv}^e}
\ev{\imath J_{jy}^b\imath J_{kz}^c\imath J_{lw}^d}&=&
-\ev{\imath J_{kz}^c\imath J_\nu}
\ev{\imath A_{mv}^e\imath\Phi_\nu\imath\Phi_\mu}
\ev{\imath J_\mu\imath J_\ga}\ev{\imath J_{jy}^b\imath J_\la}
\ev{\imath\Phi_\la\imath\Phi_\ga\imath\Phi_\de}
\ev{\imath J_\de\imath J_{lw}^d}
\nonumber\\&&
-\ev{\imath J_{kz}^c\imath J_\ga}\ev{\imath J_{jy}^b\imath J_\nu}
\ev{\imath A_{mv}^e\imath\Phi_\nu\imath\Phi_\mu}
\ev{\imath J_\mu\imath J_\la}
\ev{\imath\Phi_\la\imath\Phi_\ga\imath\Phi_\de}
\ev{\imath J_\de\imath J_{lw}^d}
\nonumber\\&&
-\ev{\imath J_{kz}^c\imath J_\ga}\ev{\imath J_{jy}^b\imath J_\la}
\ev{\imath\Phi_\la\imath\Phi_\ga\imath\Phi_\de}
\ev{\imath J_\de\imath J_\mu}
\ev{\imath A_{mv}^e\imath\Phi_\mu\imath\Phi_\nu}
\ev{\imath J_\nu\imath J_{lw}^d}
\nonumber\\&&
+\ev{\imath J_{kz}^c\imath J_\ga}\ev{\imath J_{jy}^b\imath J_\la}
\ev{\imath A_{mv}^e\imath\Phi_\la\imath\Phi_\ga\imath\Phi_\de}
\ev{\imath J_\de\imath J_{lw}^d}
\eea
gives the gluon \DS equation which in momentum space reads:
\bea
\G_{AAim}^{ae}(k)&=&
\imath\de^{ae}\left[\vec{k}^2\de_{im}-k_ik_m\right]
+\int(-\dk{\w})\G_{\ov{c}cA i}^{(0)bca}(\w-k,-\w,k)W_c^{cd}(\w)
\G_{\ov{c}cAm}^{dfe}(\w,k-\w,-k)W_c^{fb}(\w-k)
\nonumber\\&&
-\int(-\dk{\w})\G_{\phi\si Ai}^{(0)bca}(\w-k,-\w,k)W_{\si\ba}^{cd}(\w)
\G_{\ba\al Am}^{dfe}(\w,k-\w,-k)W_{\al\phi}^{fb}(\w-k)
\nonumber\\&&
-\int(-\dk{\w})\G_{A\pi\si ij}^{(0)bca}(\w-k,-\w,k)W_{\si\ba}^{cd}(\w)
\G_{\ba\al A km}^{dfe}(\w,k-\w,-k)W_{\al\pi kj}^{fb}(\w-k)
\nonumber\\&&
-\frac{1}{2}\int(-\dk{\w})\G_{3Akji}^{(0)bca}(\w-k,-\w,k)
W_{A\ba jl}^{cd}(\w)\G_{\ba\al Alnm}^{dfe}(\w,k-\w,-k)W_{\al Ank}^{fb}(\w-k)
\nonumber\\&&
-\frac{1}{6}\int(-\dk{\w})(-\dk{v})
\G_{4Alkji}^{(0)dcba}(-v,-\w,v+\w-k,k)W_{A\la jn}^{bf}(k-v-\w)
W_{A\ga ko}^{cg}(\w)W_{A\de lp}^{dh}(v)\times
\nonumber\\&&
\G_{\la\ga\de Anopm}^{fghe}(k-\w-v,\w,v,-k)
\nonumber\\&&
+\frac{1}{2}\int(-\dk{\w})\G_{4Aimlk}^{(0)aecd}(k,-k,\w,-\w)
W_{AAkl}^{cd}(-\w)
\nonumber\\&&
+\frac{1}{2}\int(-\dk{\w})(-\dk{v})\G_{4Alkji}^{(0)dcba}(-v,-\w,v+\w-k,k)
W_{A\de ln}^{df}(v)W_{A\ga ko}^{cg}(\w)\G_{\de\ga\la nop}^{fgh}(v,\w,-v-\w)
\times
\nonumber\\&&
W_{\la\mu pq}^{hi}(v+\w)\G_{\mu\nu Aqrm}^{ije}(v+\w,k-v-\w,-k)
W_{\nu Arj}^{jd}(\w+v-k).
\eea
Again, the occurence of the summation over $\al,\ldots,\la$ leads to many 
different possible loop terms.

\begin{figure*}[t]
\includegraphics[width=0.95\linewidth]{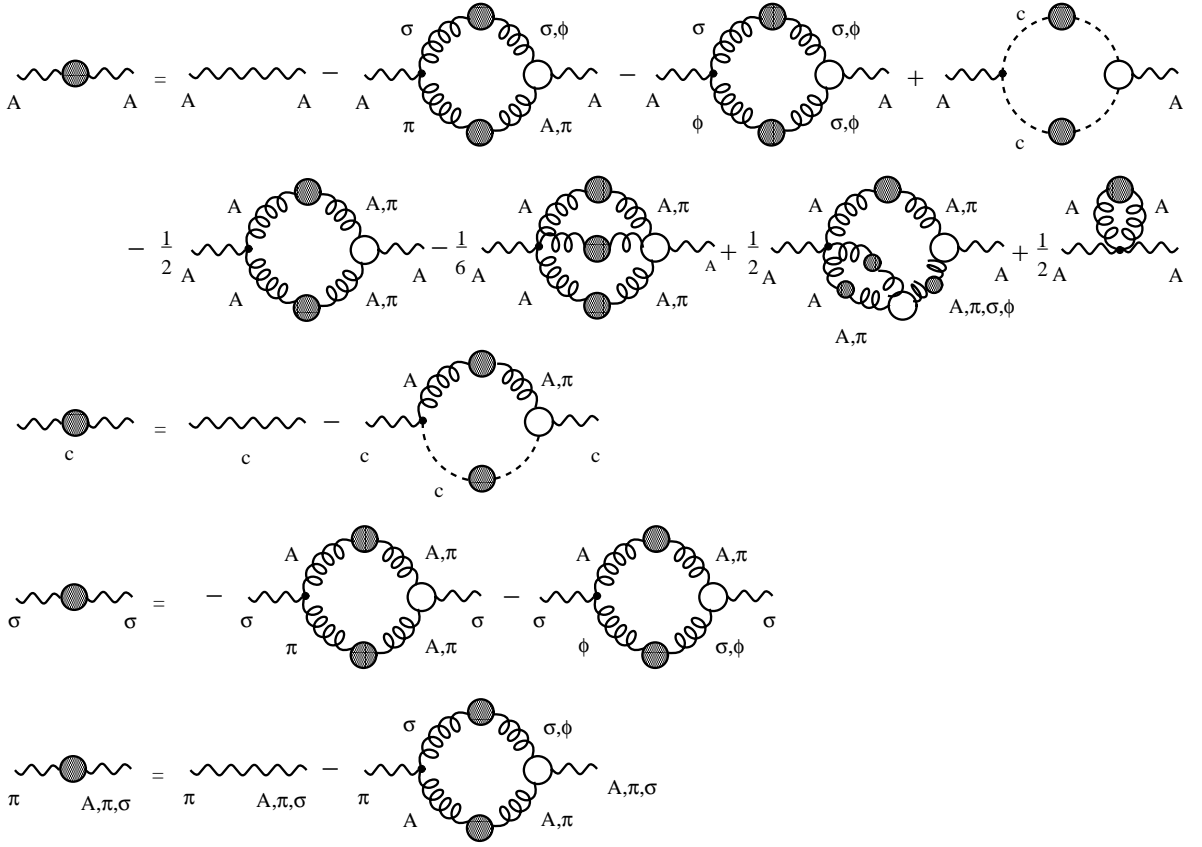}
\caption{\label{fig:dses} A diagrammatical representation of the coupled 
system of \DS equations.  Filled blobs denote dressed propagators and empty 
circles denote dressed proper vertex functions.  Wavy lines denote proper 
functions, springs denote connected (propagator) functions and dashed lines 
denote the ghost propagator.  Labels indicate the various possible 
propagator and vertex combinations that comprise the self-energy terms.}
\end{figure*}

We present the complete set of \DS equations in Figure~\ref{fig:dses}.  

\section{Summary and Outlook}

In this work, we have derived the \DS equations for Coulomb gauge Yang-Mills 
theory within the first order formalism.  In discussing the first order 
formalism it was noted that the standard BRS transform is supplemented by a 
second transform which arises from the ambiguity in setting the gauge 
transform 
properties of the $\pi$ and $\phi$ fields.  The motivation behind the use of 
the first order formalism is two-fold: the energy divergent ghost sector can 
be formally eliminated and the system can be formally reduced to physical 
degrees of freedom, formal here meaning that the resulting expressions are 
non-local and not useful for practical studies.  The cancellation of the 
ghost sector is seen within the context of the \DS equations and the Green's 
functions stemming from the local action.  It remains to be seen how the 
physical degrees of freedom emerge.

Given that the boundary conditions imposed by considering the Gribov problem 
and that the Jacobians of both the standard BRS transform and its 
supplemental transform within the first order formalism remain trivial, the 
field equations of motion and the Ward-Takahashi identity have been 
explicitly derived.  The supplemental part of the BRS transform has been 
shown to be equivalent to the equations of motion at the level of the 
functional integral and as such is more or less trivial.  Certain exact 
(i.e., not containing interaction terms) relations for the Green's functions 
of the theory have been discussed and their solutions presented.  These 
relations serve to simplify the framework considerably.  The propagators 
pertaining to vector fields are shown to be transverse, the proper functions 
involving Lagrange multiplier fields reduce to kinematical factors or vanish 
and the proper functions involving functional derivatives with respect to 
the $\phi$-field can be explicitly derived from those involving the 
corresponding $\pi$-field derivatives.

The full set of Feynman rules for the system has been derived along with the 
tree-level proper two-point functions and the general form of the two-point 
functions (connected and proper) has been discussed.  The relationship 
between the (connected) propagators and the proper two-point functions, 
stemming from the Legendre transform, has been studied.  The resulting 
equations show that within the first order formalism the dressing functions 
of the two types of two-point Green's functions are non-trivially related to 
each other.  In addition, given that there are no vertices involving 
derivatives with respect to the Lagrange multiplier fields, the set of \DS 
equations needed to study the two-point functions of the theory is reduced.

The relevant \DS equations for the system have been derived in some detail.  
It is shown how the number of self-energy terms is considerably amplified by 
the introduction of the various fields inherent to the first order 
formalism.  The \DS equations arising from the ghost fields are shown to be 
independent of the ghost energy and the vertices involving the ghost fields 
are UV finite.

Despite the complexity of dealing with a non-covariant system with many 
degrees of freedom, the outlook is positive and the rich structure of the 
\DS equations is not as intimidating as it might initially appear.  The 
non-covariance of the setting means that all the dressing functions that 
must be calculated are generally functions of two variables.  However, as 
seen from the Feynman rules, the energy dependence of the theory stems from 
the tree-level propagators alone and not from vertices (a consequence of the 
fact that the ony explicit time derivative in the action occurs within a 
kinetic term).  The time-dependence of the integral kernels will therefore 
be significantly less complicated than perhaps would otherwise occur.  Given 
the experience in Landau gauge adapting the techniques, both analytical and 
numerical, to solve the \DS equations in Coulomb gauge seems eminently 
possible though certainly challenging.  The results of such a study should 
provide a better understanding of the issues of confinement, and with the 
inclusion of quarks, the hadron spectrum.

\begin{acknowledgments}
It is a pleasure to thank R.~Alkofer and D.~Zwanziger for useful and 
inspiring discussions.  This work has been supported by the Deutsche 
Forschungsgemeinschaft (DFG) under contracts no. Re856/6-1 and Re856/6-2.
\end{acknowledgments}

\appendix
\section{\label{app:eom}Explicit form of the equations of motion}

For completeness we list the explicit equations of motion for the various 
fields represented by \eq{eq:eom}.
\bea
\ro_x^aZ[J]&=&\int\cd\Phi\left\{\s{\vec{D}_x^{ab}}{\left(\vec{\pi}_x^b
-\div_x\phi_x^b\right)}\right\}\exp{\left\{\imath\cs+\imath\cs_s\right\}},\\
J_{ix}^aZ[J]&=&-\int\cd\Phi\left\{\nabla_{ix}\la_x^a
-gf^{bac}(\nabla_{ix}\ov{c}_x^b)c_x^c-\pd_x^0\left(\pi_{ix}^a
-\nabla_{ix}\phi_x^a\right)-gf^{bac}\left(\pi_{ix}^b
-\nabla_{ix}\phi_x^b\right)\si_x^c
\right.\nonumber\\&&\left.
+\left[\de_{ij}\nabla_x^2-\nabla_{ix}\nabla_{jx}\right]A_{jx}^a
+gf^{abc}\left[A_{jx}^b\nabla_{ix}A_{jx}^c+2A_{jx}^c\nabla_{jx}A_{ix}^b
-A_{ix}^c\nabla_{jx}A_{jx}^b\right]
\right.\nonumber\\&&\left.
-\frac{1}{4}g^2f^{fbc}f^{fde}\left[\de^{ab}A_{jx}^cA_{jx}^eA_{ix}^d
+\de^{ad}A_{jx}^cA_{jx}^eA_{ix}^b+\de^{ac}A_{jx}^bA_{jx}^dA_{ix}^e
+\de^{ae}A_{jx}^bA_{jx}^dA_{ix}^c\right]
\right\}\exp{\left\{\imath\cs+\imath\cs_s\right\}},\nonumber\\\\
\et_x^aZ[J]&=&\int\cd\Phi\left\{\s{\div_x}{\vec{D}_x^{ab}}c_x^b\right\}
\exp{\left\{\imath\cs+\imath\cs_s\right\}},\label{eq:aghdse1}\\
\ov{\et}_x^aZ[J]&=&\int\cd\Phi
\left\{\s{\vec{D}_x^{ab}}{\div_x}\ov{c}_x^b\right\}
\exp{\left\{\imath\cs+\imath\cs_s\right\}},\label{eq:ghdse2}\\
\ka_x^aZ[J]&=&-\int\cd\Phi\left\{-\s{\div_x}{\vec{X}_x^a}\right\}
\exp{\left\{\imath\cs+\imath\cs_s\right\}},\label{eq:phdse1}\\
K_{ix}^aZ[J]&=&-\int\cd\Phi\left\{\nabla_{ix}\ta_x^a-X_{ix}^a\right\}
\exp{\left\{\imath\cs+\imath\cs_s\right\}},\label{eq:pidse1}\\
\xi_{\la x}^aZ[J]&=&\int\cd\Phi\left\{\s{\div_x}{\vec{A}_x^a}\right\}
\exp{\left\{\imath\cs+\imath\cs_s\right\}},\label{eq:ladse1}\\
\xi_{\ta x}^aZ[J]&=&\int\cd\Phi\left\{\s{\div_x}{\vec{\pi}_x^a}\right\}
\exp{\left\{\imath\cs+\imath\cs_s\right\}},\label{eq:tadse1}
\eea
with $X$ defined in \eq{eq:X}.

\section{\label{app:jac}Jacobian Factors}
\setcounter{equation}{0}
The BRS and $\al$-transforms, \eq{eq:brstrans} and \eq{eq:altrans} 
respectively, can be regarded as changes of variable in the functional 
integral and that the action is invariant will lead to Ward-Takahashi 
identities.  However, as with all changes of integration variable, we must 
consider the relevant Jacobian factors.  These turn out to be trivial, 
although this is not immediately obvious.  Starting with the BRS transform, 
we begin with the observation that $\de\la^2=0$ (since $\de\la$ is 
Grassmann-valued) which means that in the Jacobian determinant only the 
diagonal elements will contribute since all off-diagonal elements are 
${\cal O}(\de\la)$.  Besides the trivial unit terms of the form 
$\de^{ab}\de(x-y)$ (with an extra $\de_{ij}$ factor for vector fields) all 
diagonal terms have the color structure $f^{abc}H^c$, as can be seen in the 
form of the transform.  Given that 
\be
\mathrm{Det}(\de^{ab}+f^{abc}H^c)
=\exp{\left\{\mathrm{Tr}\mathrm{Log}(\de^{ab}+f^{abc}H^c)\right\}}
\ee
only the first term proportional to $f^{abc}$ survives when the logarithm is 
expanded ($H\sim\de\la$).  We then see that $\mathrm{Tr}f^{abc}H^c=0$ which 
leaves only the unit term of the exponential.  The Jacobian for the BRS 
transform is thus trivial.

To see that the Jacobian involved for the $\al$-transform is trivial is 
slightly more involved.  First we note that the only non-trivial part of the 
matrix of variations involves only the $\vec{\pi}$ and $\phi$ fields, all 
other rows or columns reducing to trivial identity contributions.  The 
sub-matrix of variations for the $\vec{\pi}$ and $\phi$ fields can be 
written $\openone+K$ where $K\sim f^{abc}\th_x^c$ and is independent of the 
fields.  Since the transform is a change of variables, the functional 
integral is independent of $\th_x^c$ and we can write
\be
0=\left.\frac{\de Z}{\de\th_x^c}\right|_{\th=0}
=\frac{\de}{\de\th_x^c}\left.J[\th]\int\cd\Phi
\exp{\left\{\imath\cs+\imath\cs_s[\th]\right\}}\right|_{\th=0}.
\ee
With this in mind, it suffices to show that
\be
\left.\frac{\de J[\th]}{\de\th_x^c}\right|_{\th=0}=0
\ee
in order for the actual form of the Jacobian to be irrelevant (note that 
$J[\th=0]=1$).  We can write
\be
J[\th]=\exp{\{\mathrm{Tr}\mathrm{Log}(1+K[\th])\}}
\ee
and so
\be
\left.\frac{\de J[\th]}{\de\th_x^c}\right|_{\th=0}
=\left.\mathrm{Tr}\left[\frac{\de K[\th]}{\de\th_x^c}
-K[\th]\frac{\de K[\th]}{\de\th_x^c}+\ldots\right]\right|_{\th=0}J[\th=0].
\ee
Since $K$ is linear in $\th$, only the first term of the expansion is 
present when $\th=0$.  This term has the color structure $f^{abc}$, which 
vanishes under the trace operation and so, indeed
\be
\left.\frac{\de J[\th]}{\de\th_x^c}\right|_{\th=0}=0
\ee
and the Jacobian, although itself not unity is trivial and does not further 
enter the discussion of the functional integral under the transform.


\end{document}